\documentclass[structabstract]{aa}  
\usepackage{graphicx}
\usepackage{txfonts}
\usepackage{natbib}
\usepackage{color}
\begin{document}

   \title{Low-resolution spectroscopy of main sequence stars\\belonging to 12
   Galactic globular clusters.\\I. CH and CN band strength
   variations.\thanks{Based on FORS observations collected at the
   European Southern Observatory, Chile, within the observing programs 68.D-0510
   and 69.D-0056. Also based on data obtained from the ESO Archive, within the
   observing program 67.D-0153.}}

\titlerunning{CN and CH band strengths of MS stars in 12 GGC} 

   \author{E. Pancino \inst{1},  M. Rejkuba\inst{2}, M. Zoccali\inst{3} \and R.
	   Carrera\inst{1,4},}
   \authorrunning{Pancino et al.}
   \institute{INAF-Osservatorio Astronomico di Bologna, via Ranzani 1, 40127, Bologna,
             Italy\\
             \email{elena.pancino@oabo.inaf.it}
             \and
	ESO, Karl-Schwarzschild-Strasse 2, 85748 Garching, Germany
             \and
	Universidad Cat\'olica de Chile, Departamento de Astronom\'{\i}a y
             Astrof\'{\i}sica, Casilla 306, Santiago 22, Chile
	     \and
	Instituto de Astrof\'{\i}sica de Canarias, E-38205 La Laguna, Tenerife, 
	     Spain     
             }

   \date{Received September XX, XXXX; accepted March XX, XXXX}

 
  \abstract
   {Globular clusters show star-to-star abundance variations for light elements
    that are not yet well understood. The preferred explanation involves a {\em
    self-enrichment scenario}, within which two subsequent generations of stars
    co-exist in globular clusters. Observations of chemical abundances in the main
    sequence and sub-giant branch stars allow us to investigate the signature of
    this chemically processed material without the complicating effects caused by
    stellar evolution and internal mixing.}
   {Our main goal is to investigate the carbon-nitrogen anti-correlation with
    low-resolution spectroscopy of 20--50 stars fainter than the first dredge-up
    in seven Galactic globular clusters (NGC~288, NGC~1851, NGC~5927, NGC~6352,
    NGC~6388, and Pal~12) with different properties. We complemented our 
    observations with 47~Tuc archival data, with four additional clusters from the
    literature (M~15, M~22, M~55, NGC~362), and with additional literature data on
    NGC~288.}
   {In this first paper, we measured the strengh of the CN and CH band
   indices, which correlate with the N and C abundances, and we investigated
   the anti-correlation and bimodality of these indices. We compared r$_{CN}$,
   the ratio of stars belonging to the CN-strong and weak groups, with 15 
   different cluster parameters.}
   {We clearly see bimodal anti-correlation of the CH and CN band stregths
   in the metal-rich clusters (Pal~12, 47~Tuc, NGC~6352, NGC~5927). Only M~15
   among the metal-poor clusters shows a clearly bimodal anti-correlation. We
   found weak correlations (sligthly above 1\,$\sigma$) of r$_{CN}$ with the
   cluster orbital parameters, present-day total mass, cluster concentration,
   and age.}
   {Our findings support the {\em self-enrichment scenario}, and suggest that
   the occurrence of more than two major generations of stars in a GGC should be
   rare. Small additional generations ($<$10--20\% of the total) would be
   difficult to detect with our samples. The first generation, which corresponds
   to the CN-weak stars, usually contains more stars than the second one
   ($<$r$_{CN}$$>$=0.82$\pm$0.29), as opposed to results based on the
   Na-O anti-correlations.}

   \keywords{stars: abundances -- stars: main sequence --globular clusters:
   individual (M~15, M~22, M~55, 47~Tuc, NGC~288, NGC~362, NGC~1851,
   NGC~5927, NGC~6352, NGC~6388, NGC~6752, and Pal 12)}

   \maketitle
%

\section{Introduction}
\label{sec-intro}

Recent work revealed that several Galactic globular clusters (hereafter GGC) had a
much more complex star-formation history than previously thought. This has
implications for their use as simple stellar population (SSP) templates to study
more complex and distant stellar systems and galaxies, and therefore requires
careful study. 

Based on deep and accurate photometric studies, GGC were found to contain
different stellar populations, in the form of multiple evolutionary sequences in
their color-magnitude diagrams (CMD). The first cluster found to host multiple
stellar populations was $\omega$ Cen
\citep{lee99,pancino00,bedin04,sollima05,villanova07}. Lately NGC 2808
\citep{piotto07}, NGC 1851 \citep{milone08,ventura09b,han09}, NGC 6388
\citep{moretti09}, M~22 \citep{piotto09}, and 47 Tuc \citep{anderson09} joined the
rapidly growing group of complex stellar population GGC.

\begin{table*}
\caption{Observing logs (see Sect.~\ref{sec-red} and \ref{sec-sample} for
details).}
\label{logs}
\begin{center}
\begin{tabular}{l r r r r r r r r r r r} 
\hline \hline
Cluster & [Fe/H] &(m--M)$_V$ & E(B--V) & run & n$_{stars}$ & n$_{fields}$ & t$_{exp}^{Tot}$ & S/N(3800\AA) \\
        & (dex)  & (mag)   & (mag)     &     &             &              & (sec)           &              \\
\hline
NGC 104 (47~Tuc) & --0.68 & 13.37 & 0.04 & 2001-07 (Archive) & 64 & 3 & 11700 & 20.5$\pm$7.8 \\
NGC 288          & --1.39 & 14.83 & 0.03 & 2001-10           & 72 & 5 & 19800 & 23.6$\pm$4.8 \\
NGC 1851         & --1.22 & 15.47 & 0.02 & 2001-10           & 50 & 3 & 18000 & 28.7$\pm$3.6 \\
NGC 5927         & --0.55 & 15.81 & 0.45 & 2002-07	     & 45 & 4 & 32400 &  9.9$\pm$1.9 \\
NGC 6352         & --0.54 & 14.44 & 0.21 & 2002-07	     & 24 & 1 &  9900 & 19.0$\pm$3.0 \\
NGC 6388         & --0.44 & 16.14 & 0.37 & 2002-07	     & 22 & 2 & 19800 &  8.9$\pm$2.5 \\
NGC 6752         & --1.50 & 13.13 & 0.04 & 2001-10           & 56 & 4 &  7200 & 28.0$\pm$5.7 \\
Pal 12           & --0.80 & 16.47 & 0.02 & 2001-10           & 23 & 2 &  9900 & 14.1$\pm$3.9 \\
\hline\hline	  
\end{tabular}
\end{center}
\end{table*}

But even before the multiple photometric sequences were detected, 
anti-correlations between the strength of the CN and CH bands around 3880, 4200,
and 4300~\AA\  were found, starting with the pioneering work of
\citet{osborn1971}, in all properly observed clusters. These anti-correlations are
sometimes bimodal in nature (see Sect.~\ref{sec-bimo} for references), with two
separate groups of CN-strong and CN-weak stars. The first detections were of
course limited to the red giant branch (RGB) stars \citep[see][for early
reviews]{smith87,kraft94}, because of the instrumental limitations at that time,
but later they were also found among sub giant branch (SGB) stars and even among
main sequence (MS) stars \citep{cannon1998,cohen99,harbeck03,carretta05}.
Anti-correlations were found among other light elements such as Na, O, Al, Mg
\citep[see the reviews by][]{kraft94,gratton04} and F \citep{cunha03,smith05}.

No sample of Milky Way field stars observed so far showed any sign of
anti-correlations \citep[but they do show signs of dredged up CNO processing,
see][]{hanson98,gratton00,mishenina06}. Also, no chemical anomalies were found in
open clusters \citep[see references in][]{desilva09,panci09} or in the field stars
belonging to dwarf galaxies in the Local Group. Only recently, Na-O
anti-correlations were detected in extragalactic GC (globular clusters) of the
Fornax dwarf galaxy and of the Large Magellanic Clouds \citep{let06,joh06,muc09}.
{\em This suggests that environment plays a fundamental r\^ole in the occurrence
of these anomalies.} 

The presence of CH-CN anti-correlations was soon interpreted as a signature of
CN(O) cycle processing, which tends to deplete C and to enhance N, and of mixing
that can bring CN(O) processed material to the stellar surface, during the first
dredge-up that happens at the base of the RGB. To reach this conclusion, it is
assumed that CN traces the N abundace and CH the C abundance \citep{smith96}. This
is well supported by abundance analysis work based on spectral synthesis, which
revealed [C/Fe] and [N/Fe] spreads of up to more than 1~dex  \citep[see
e.g.,][]{briley04a,briley04b,cohen05}. However, this {\em intrinsic} scenario,
often referred to as the {\em stellar evolution scenario}, was soon put into
trouble by two observational facts: {\em (i)} CH and CN anti-correlations in MS
stars cannot be produced by the stars themselves, because they have low mass
(M$\simeq$0.8~M$_{\odot}$) in GGC and do not burn hydrogen through the CNO cycle;
{\em (ii)} the presence of Na, Al, Mg, and O (anti-)correlations points towards
the NeNa and MgAl chains that take place at much higher temperatures 
\citep{denisenkov90,langer93,prantzos07}, which are not reached in the MS or RGB
phases of GGC stars. The addition of internal pollution mechanisms such as
rotation-induced mixing \citep[e.g.][]{sweigart79,charbonnel95} or the so-called
canonical extra-mixing \citep[e.g.][]{denissenkov03a} did not solve the problem
because they still could not explain the Na-O and Al-Mg anti-correlations. Of
course, mixing does occur among red giants above the base of the RGB
\citep{smith96,gratton00,smith03}, and this is the reason why it is better to
focus on stars on the MS and SGB, which still have not undergone any mixing
episode, to have a clean perspective on chemical anomalies, independent from
stellar evolution effects.

A natural alternative was then some sort of {\em extrinsic} scenario, where the
chemical anomalies are produced outside the GGC stars. An example is the {\em
primordial scenario} proposed by \citet{cannon1998}, which foresaw the presence of
two clouds with different chemical composition, merging in the formation phase of
GGC. Another example is the {\em stellar pollution scenario}
\citep{dantona83,thoul02}, where only the stellar surfaces are polluted by CN(O)
processed winds. These and other {\em extrinsic} scenarios are extensively
discussed by \citet{harbeck03}, and \citet{kayser08}.  

The most promising of the {\em extrinsic scenarios} remains the so-called {\em
self-enrichment scenario}, first proposed by \citet{cottrell81}, which assumes the
presence of two (or more) subsequent stellar generations, the first polluting the
intra-cluster medium from which the second one forms. A very important constraint
to this scenario is that most GGC show a remarkably homogeneous composition as far
as heavier elements (e.g., Ca, Fe) are concerned. In the past five years, the
chase has thus moved from finding the most promising scenario to seeking for the
first generation stars responsible for polluting the intra-cluster medium.  Three
proposals have been made so far: {\em (i)} intermediate mass (M$>$3--5
M$_{\odot}$) asymptotic giant branch (AGB) stars
\citep[e.g.][]{ventura01,denissenkovherwig03,ventura09}, which naturally explain
chemical anomalies, but also provide clear links to the helium problem, proposed
to explain the multiple populations found in some clusters (see above) and their
horizontal branch (HB) morphologies \citep[see the series of papers by Carretta et
al., the last ones being][]{car09a,car09b}; {\em (ii)} fast rotating massive stars
\citep{maeder06,prantzos06,decressin07}, which suffer from meridional circulation,
able to bring CN(O) processed material to the surface, and to release it through
slow winds, through their accretion disc and, finally, through explosion; this
scenario helps also in solving the helium and multiple population problems; and
{\em (iii)} the very recently proposed massive interactive binaries
\citep{demink09}, which are known to lose mass and could indeed explain the
observed anomalies. Moreover, the slow winds of all these types of polluters are
prone to different degrees of stripping depending on the cluster and its
environment. This can explain why different clusters show different behaviours of
their multiple populations and chemical anomalies, as described in detail in
Sect.~\ref{sec-trends}. 

The aim of this paper is to look for CH and CN band strength variations and
bimodalities in a sample of unevolved stars in 12 GGC with different properties.
Those clusters that show clear variations and bimodalities will be further studied
in a following paper dedicated to abundance determinations, to investigate the
extent of the [C/Fe] and [N/Fe] variations. Main sequence stars have not undergone
any dredge-up phenomenon yet, so they are free from any evolutionary effect that,
superposed to the CH and CN {\em extrinsic} variations, could confuse the picture.
We will also correlate the CH and CN strength variations with GGC parameters, in
order to provide additional observational constraints to the {\em self-enrichment
scenario}.

\begin{figure*}
\centering
\includegraphics[angle=270,width=\textwidth]{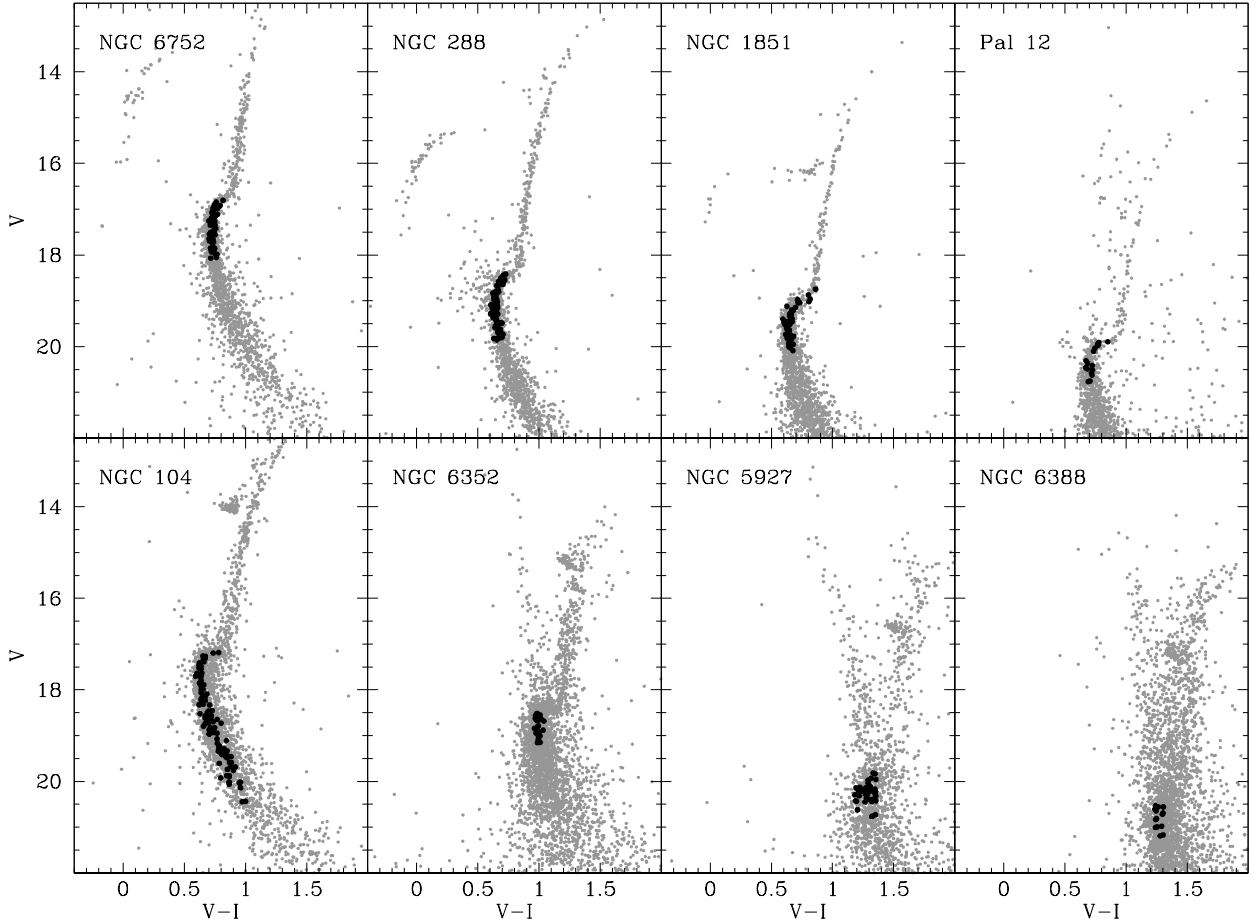}
\vspace{-1cm}
\caption{Colour-magnitude diagrams for the newly observed GGC in our sample. Grey
dots show our pre-imaging V, V--I photometry, calibrated with literature data (see
text). For 47 Tuc, which was not part of our original observations, we used the
photometry by \citet{rosenberg00}. A selection on the photometric parameters from
DAOPHOT II is applied to show only the best measured stars. Black dots mark
spectroscopic targets. Clusters are ordered as a function of their metallicity
from the most metal-poor (top left) to the most metal-rich (bottom right).}
\label{fig_dcm}
\end{figure*}

The plan of the paper is the following: in Sect.~\ref{sec-red} we discuss the
observations and data reductions; in Sect.~\ref{sec-index} we define the indices
passbands and their measurement; in Sect.~\ref{sec-res} we present our results;
in Sect.~\ref{sec-sample} we discuss the results on a cluster by cluster basis;
in Sect.~\ref{sec-trends} we correlate the CH and CN strengths with cluster
parameters, and in Sect.~\ref{sec-concl} we summarize our findings and draw our
conclusions.

\section{Observations and data reduction}
\label{sec-red}

We selected a sample of eight GGC with different properties to test the
correlations of their CH and CN strengths with cluster parameters and
environmental conditions: 47~Tuc (from the ESO archive), NGC~1851, NGC~288,
NGC~5927, NGC~6352, NGC~6388, NGC~6752, and Palomar~12. We also tried to select
stars covering a large area in each cluster, to be free from biasses due to
possible radial trends. Measurements for four more GGC were added from
\citet{kayser08}: M~15, M~22, M~55, and NGC~362, plus measurements for additional
stars in NGC~288 (see Sect.~\ref{sec-kayser}). A more detailed discussion of each
cluster's properties and results can be found in Sect.~\ref{sec-sample}.

\subsection{Observations}

Observations were carried out mostly in service mode in two different runs and two
pre-imaging runs at the ESO VLT/UT2 telescope in Cerro Paranal, Chile, with the
FORS2 multi-object spectrograph \citep{fors} in MXU mode. The observing logs can
be found in Table~\ref{logs}. The first run was dedicated to the three metal-poor
clusters (NGC~6752, NGC~288, and NGC~1851) and to Pal~12, and was carried out
between September and October 2001, with relatively good conditions for the
pre-imaging (sky mostly clear and seeing $\simeq$1") and less favorable conditions
for the spectroscopy observations (sky mostly veiled and seeing ranging from 0.5"
to 2"). The second run was dedicated to the metal-rich clusters (NGC~6352,
NGC~5927, and NGC~6388) and was carried out between April and July 2002. The
conditions for both pre-imaging and spectroscopy observations where less fortunate
than the in the first run, with the sky often veiled or cloudy, and seeing between
1.5"--2". This explains in part why the colour-magnitude diagrams (CMD) of these
clusters are of a much worse quality (Fig.~\ref{fig_dcm}) and why the S/N ratios
of their spectra tend to be quite low as well (Table~\ref{logs}). 

Although we used the same grism, GRIS\_600B, the old 2k$\times$2k SITE detector
(24 $\mu$m/pix) was replaced by a mosaic composed by two 2k$\times$4k MIT CCDs (15
$\mu$m/pix) between the two runs. This implies that the covered wavelength ranges
were not exactly the same in the two runs; also, the red sensitivity of the CCD
was improved, but this had no effect on our measurements. The typical resolution
of our spectra in both runs is $R$=$\lambda / \delta \lambda$$\simeq$800.

\subsection{Pre-imaging reductions}

We obtained a set of images for one or more fields within each cluster, generally
consisting of a short exposure (a few seconds) and a longer one (a few tens of
seconds) for each of the Johnson V and the Cousins I filters. All images were
bias-subtracted and flat-fielded with the ESO FORS pipeline. Point spread function
(PSF) fitting photometry  was carried out with the DAOPHOT~II, ALLSTAR  and
ALLFRAME  packages \citep{daophot,allframe} using a constant model PSF across the
field, which experience showed to yield the best results. 

The photometric calibration was done using stars in common with photometry from
the literature: we used the V, I catalogue published by \citet{bellazzini01} for
NGC~1851; the B, V catalogue by \citet{dalessandro08} supplemented by I magnitudes
(Dalessandro, private communication) for NGC~6388; and the V, I catalogue by
\citet{rosenberg00} for 47~Tuc (NGC~104). For all remaining clusters we used the
HST photometries by \citet{piotto02}. The resulting calibrated colour-magnitude
diagrams and programme stars are shown in Fig.~\ref{fig_dcm}.

\subsection{Spectroscopy reductions}

The spectroscopic target stars were selected from the pre-imaging data as the most
isolated stars located around the turn-off (TO) and SGB regions of the CMD
(Fig.~\ref{fig_dcm}). In most cases, more than one exposure was taken with a
single mask, to reach a higher S/N ratio. With each mask, we were able to observe
20--30 stars with 1" wide, 10" long slitlets. The average S/N for each cluster
(after the frame pre-reductions) in the CN 3880~\AA\  region, together with the
total number of good stars analysed in each cluster (see Sect.~\ref{sec-1stqc}),
is indicated in the last two columns of Table~\ref{logs}. 

For the data pre-reduction, we used IRAF\footnote{IRAF is distributed by the
National Optical Astronomy Observatory, which is operated by the Association of
Universities for  Research in Astronomy, Inc., under cooperative agreement with
the National Science Foundation.} for overscan correction and bias-subtraction.
We removed cosmic ray hits with the IRAF Laplacian edge-detection routine
\citep{dokkun01}. The frames were subsequently flat-fielded and reduced to
one-dimension spectra with the IRAF task {\em apall}. The sky background was
subtracted using the information on both sides of the stellar spectrum. Finally,
for those fields that were observed more than once, we co-added all spectra
together on a star by star basis. We checked that the shift between the
different spectra ($\leq$10~km~s$^{-1}$) was negligible compared to our
wavelength calibration uncertainty ($\simeq$40~km~s$^{-1}$) before coadding. 

\begin{figure}
\centering
\includegraphics[bb=30 150 590 430,width=\columnwidth]{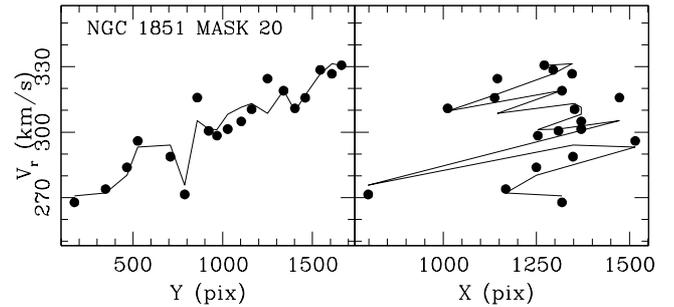}
\caption{Influence of telescope and/or instrument flexures on the derived radial
velocities. Each point represents one star, with its radial velocity and the X
and Y coordinates of the corresponding slitlet center. The solid lines represent
a bilineal fit to the data (see text).}
\label{fig_flexures}
\end{figure}

The shape of the final spectra is affected by the different response of the
instrument at different wavelengths. There are two ways to remove this effect:
flux calibration and continuum normalization. The first needs observations of
standard stars, which were not available in our case. The alternative is to fit a
polynomial to the pseudo-continuum and use it to normalize the spectrum. However,
the CN 3880~\AA\   band region contains many atomic lines and molecular bands, and
the response of the instrument is quite low in this ultraviolet region. Therefore,
no attempt was made to normalize the spectra, as many other authors skip this step
for the same reason \citep[e.g.][among others]{cohen02,harbeck03,kayser08}.

\subsection{Radial velocity and shifts}

Finally, radial velocities were calculated by cross-correlation (with {\em fxcor}
in IRAF), choosing the highest S/N star on each MXU plate as a template. The
templates $V_r$ were computed using the laboratory positions of the strongest
lines (e.g. H$_{\beta}$, H$_{\delta}$, Ca H and K) in our wavelength region with
the IRAF task {\em rvidlines}. 

The three dominant sources of uncertainty for the final $V_r$ of our spectra are:
(i) uncertainty in the wavelength calibration; (ii) small shifts in the star
centering within each slitlet; and (iii) variations across the field of view owing
to instrument flexures. The uncertainties in the wavelength calibration are
generally removed using the positions of strong emission sky lines, which could
also be used to evaluate the influence of the instrument flexures
\citep[e.g.][]{gallart01}. Unfortunately, in the wavelength region observed here,
there are no strong sky lines that can be used. The uncertainty due to small
misalignment of stars in the slitlets could be estimated with the through-slit
images \citep[see][]{carrera07}, and we found it negligible ($<$10~km~s$^{-1}$) in
comparison with the $V_r$ errors ($\sim$20--40~km~s$^{-1}$, see below). 

\begin{table*}
\caption{Index measurements and radial velocities for the sample stars. A
portion of the table is shown for guidance about its content, the complete table
is available in electronic format through the ADS service.}
\label{electronic}
\begin{center}
\begin{tabular}{l l c c c c c c c c c c} 
\hline \hline
Cluster & Star & RA & Dec & CN & $\delta$CN & err$_{CN}$ & CH & $\delta$CH
& err$_{CH}$ & $V_r$ & $\sigma_{V_r}$ \\
& & (deg) & (deg) & (mag) & (mag) & (mag) & (mag) & (mag) & (mag) & 
(km~s$^{-1}$) & (km~s$^{-1}$) \\
\hline
NGC~104 & 0101 & 6.58283 & -71.88794 & -0.262 & -0.266 & 0.029 & 0.837 &  0.872 & 0.040 & -52 & 33 \\
NGC~104 & 0102 & 6.58437 & -71.88708 & -0.313 & -0.236 & 0.035 & 1.024 &  0.953 & 0.058 & -23 & 40 \\
NGC~104 & 0104 & 6.59971 & -71.88342 & -0.290 & -0.147 & 0.059 & 1.020 &  1.016 & 0.085 & -40 & 52 \\
NGC~104 & 0105 & 6.63296 & -71.88597 & -0.404 & -0.128 & 0.061 & 1.077 &  1.024 & 0.106 & -36 & 49 \\
NGC~104 & 0106 & 6.65038 & -71.88544 &  0.074 & -0.184 & 0.060 & 0.978 &  0.987 & 0.069 & -44 & 45 \\
NGC~104 & 0107 & 6.58404 & -71.87014 & -0.454 & -0.310 & 0.018 & 0.902 &  0.877 & 0.042 & -12 & 28 \\
NGC~104 & 0109 & 6.64762 & -71.87739 & -0.393 & -0.247 & 0.033 & 1.000 &  0.938 & 0.064 & -19 & 31 \\
\hline\hline
\end{tabular}
\end{center}
\end{table*}

Significant variations of $V_r$ across the field of view, most probably due to
flexures, are clearly observed as shown in Fig.~\ref{fig_flexures}. The estimated
radial velocity varies systematically with the position of each slitlet along the
Y CCD axis (and on the X axis as well), by as much as $\sim$80 km~s$^{-1}$, i.e.,
larger than the the velocity uncentainty ($\sim$20--40 km~s$^{-1}$) and of the
scatter expected within each cluster (of the order of a few km s$^{-1}$). The
example in Fig.~\ref{fig_flexures} shows a typical difference of about $\simeq$60
km s$^{-1}$. To correct for these variations, we computed a bilineal fit for each
observed plate, in the form $V_r$$=$$a+bX+cY$ (see Fig.~\ref{fig_flexures}). Stars
near the centre of each MXU mask, which have the smallest shifts, were used to
roughly report all the mask velocities to the GGC systemic velocity
\citep[from][]{harris96}. The final velocities corrected for the above variations
are reported in Table~\ref{electronic}. 

\subsection{Quality control and sample selection}
\label{sec-1stqc}

Given the huge amount of spectra (see Table~\ref{logs}) and the sometimes low S/N
achieved during observations, we applied the following selection criteria:

\begin{itemize}
\item{spectra with S/N$<$8 (per pixel) in the CN 3880~\AA\  band region were
rejected;}
\item{spectra with significant defects (spikes, holes) in the measurement
windows were rejected;}
\item{stars with discrepant radial velocity by more than 2.5\,$\sigma$ from the
cluster average were rejected;}
\item{stars with discrepant Ca(H+K) index measurement were rejected (see
Sect.~\ref{sec-index} for details);}
\item{stars with discrepant H$_{\beta}$ index measurements were rejected (see
Sect.~\ref{sec-index} for details);}
\end{itemize}

The criterion that had the highest impact on star rejections was by far the
first one. The final sample contains 356 stars (out of an initial set of
$\sim$600) belonging to eight clusters, as indicated in the third column of
Table~\ref{logs} and also reported in Table~\ref{electronic}.

\begin{figure}
\centering
\includegraphics[bb=60 160 580 700,width=\columnwidth]{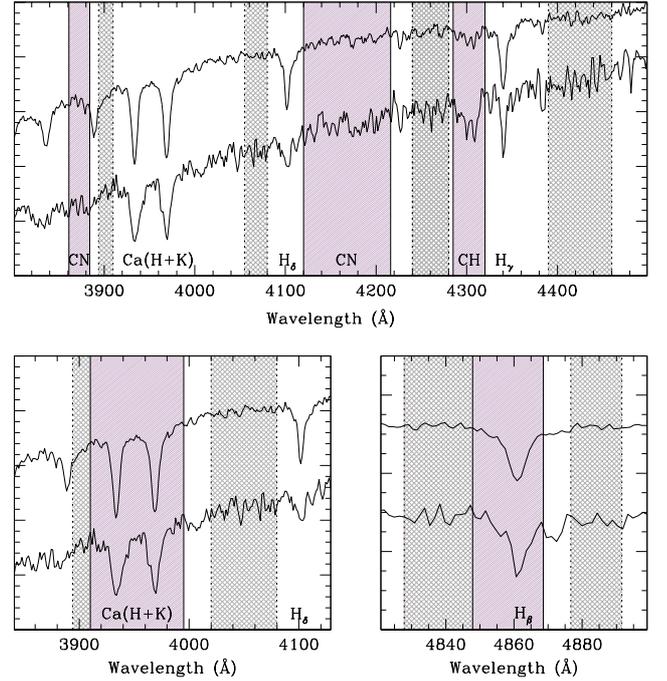}
\caption{Two spectra of CN-weak, CH-strong stars with different S/N ratio
are shown as examples of our best and worst data: the upper spectrum in all
panels has S/N$\simeq$25 (on the S3839(CN) band) and belongs to NGC~1851; the
lower spectrum has S/H$\simeq$10 and belongs to NGC~6388. Both are shifted on an
arbitrary flux scale for sake of clarity. The top panel illustrates the chosen
windows used for the CN and CH indices (magenta hatched regions) along with
their respective continuum windows (grey hatched regions); while the bottom
panels show the windows adopted for the H and K Calcium index (bottom left) and
the H$_{\beta}$ index (bottom right).}
\label{fig_spindex}
\end{figure}

\section{Index definition and measurement}
\label{sec-index}

The strength of a spectral feature can be evaluated with a {\em spectral index},
which is defined through one window centred on the atomic line or molecular band
that we will evaluate, and one or more windows around it, used to define the
continuum level. In the literature we can find many definitions of the spectral
indices used to measure the strength of the CN and CH molecular bands. In most
cases, these definitions are optimized for red giant stars, observed at a specific
spectral resolution. In our case, because we observe subgiants and dwarfs, and we
aim to compare our results with others in the literature, we chose to adopt the
indices defined by \cite{harbeck03}:

\begin{eqnarray*}\label{indices}
     S3839(CN)&=&-2.5\log \frac{ F_{3861-3884}}{ F_{3894-3910}}\\
     S4142(CN)&=&-2.5\log \frac{ F_{4120-4216}}{ 0.5 F_{4055-4080}+0.5 F_{4240-4280}}\\
     CH4300&=&-2.5\log \frac{ F_{4285-4315}}{0.5 F_{4240-4280}+0.5 F_{4390-4460}},\\
\end{eqnarray*}

\noindent where $F_{3861-3884}$, for example, is the summed spectral flux in ADU
counts from 3861 to 3884 \AA. The related uncertainties have been obtained with
the expression derived by \citet{vollmann06}, assuming pure photon (Poisson)
noise statistics in the flux measurements\footnote{We could neglect here the
readout noise, always between 3 and 5 ADU, compared to our 300--3000 ADU
gathered in the CN band region; the error in the sky subtraction, with tens of
pixels on each side of the spectrum for sky estimation, and an order of 100 ADU
sky level for the worst cases; and the spectrum tracing errors, because we always
had a peak signal -- in the central pixel -- of at least 70 ADU, and a total
300-3000 ADU level, even in the CN region that had the lowest S/N. }.
Fig.~\ref{fig_spindex} (top panel) shows the adopted indices windows; examples
of our best and worst S/N spectra are overplotted.

We also measured two additional indices, centred around the calcium H and K
lines and the H$_{\beta}$ line (see also Fig.~\ref{fig_spindex}, bottom
panels): 

\begin{eqnarray*}\label{more-indices}
     HK&=&1-\frac{F_{3910-4020}}{ F_{4020-4130}}\\
     H_{\beta}&=&1-\frac{F_{4847.875-4876.625}}{0.5F_{4827.875-4847.875}+0.5F_{4876.625-4891.625}},
\end{eqnarray*}

\noindent The H and K line strengths depend mostly on temperature and on the
calcium abundance, while the H$_{\beta}$ line strength depends on temperature
and of course on the hydrogen abundance. We used these two indices to reject a
few remaining 3\,$\sigma$ outliers (at most 2--3 stars per cluster). Since we
used the indices only in a relative sense, we did not need to compare our
measurements with the literature, therefore we defined our own narrow and
conservative windows (see Fig.~\ref{fig_spindex}, lower panels) in order to
minimize the spread due to disturbing features containing other elements and to
maximize our ability to pinpoint outliers.

\subsection{CN-band choice}

\begin{figure}
\centering
\includegraphics[width=\columnwidth]{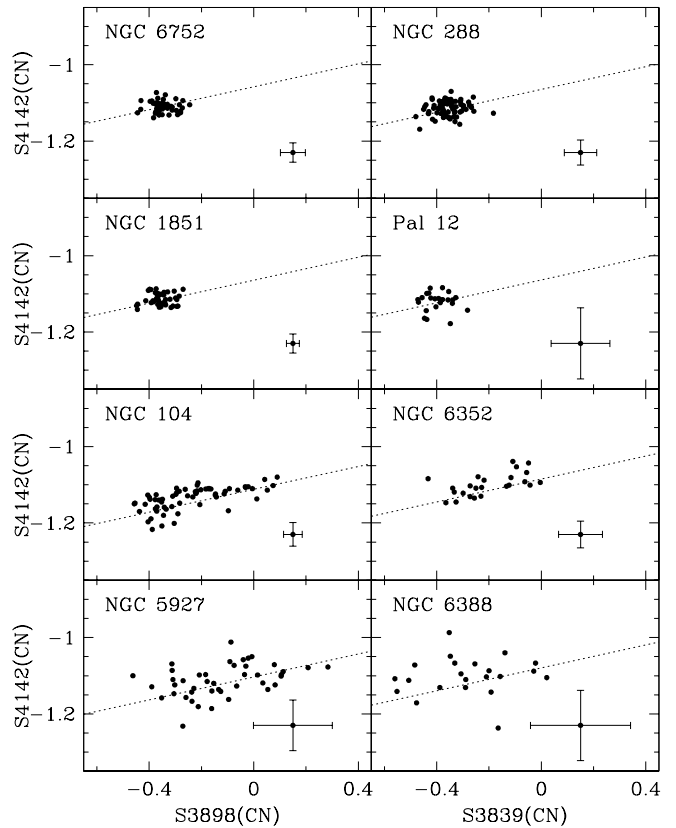}
\caption{Comparison between the CN S3839 and the S4142 indices. Each panel shows
measurements for the indicated cluster, and the median errobar is also shown in
the lower right corner. Clusters are sorted by metallicity, from the most
metal-poor (top-left) to the most metal-rich (bottom-right). Dotted lines are
linear fits with a fixed slope (that of the whole sample) and rougly adjusted
zeropoints.}
\label{fig_cncn2}
\end{figure}

In our wavelength region, we measured two different CN band indices, S3839 for the
CN band around 3880\AA\   and S4142 for the much weaker one around 4200 \AA\ 
\citep{harbeck03}. Fig.~\ref{fig_cncn2} shows a comparison between the two. The
correlation appears weak for the metal-poor clusters (NGC~6752, NGC~288,
NGC~1851), in spite of the higher S/N of the spectra, because these double-metal
molecular bands are weaker. A clear correlation appears for the most metal-rich
clusters (47~Tuc, NGC~6352, NGC~5927, and NGC~6388), although it is not as
striking for NGC~6388, given the low S/N and paucity of stars. The slope of the
relation between S4142 and S3839 appears to be roughly the same for all clusters
(dotted line in Fig.~\ref{fig_cncn2}), while the zeropoint varies slightly from
cluster to cluster. 

As expected, Fig.~\ref{fig_cncn2} clearly shows that S4142 is less sensitive to
the CN variations, because generally its spread is just slightly larger than the
median errorbar shown in the lower-right corner of each panel. This is especially
true for the metal-poor clusters. On the other hand, the spread of S3839 is always
larger than its median errobar, even for the metal-poor clusters. For this reason,
we will be relying on S3839 measurements for CN, and we will set aside the less
reliable S4142 ones \citep[as done also by, e.g.,][]{cannon1998,harbeck03}.

\subsection{Dependency on temperature and gravity}
\label{sec-tg}

\begin{figure}
\centering
\vspace{0.3cm}
\includegraphics[angle=270,width=\columnwidth]{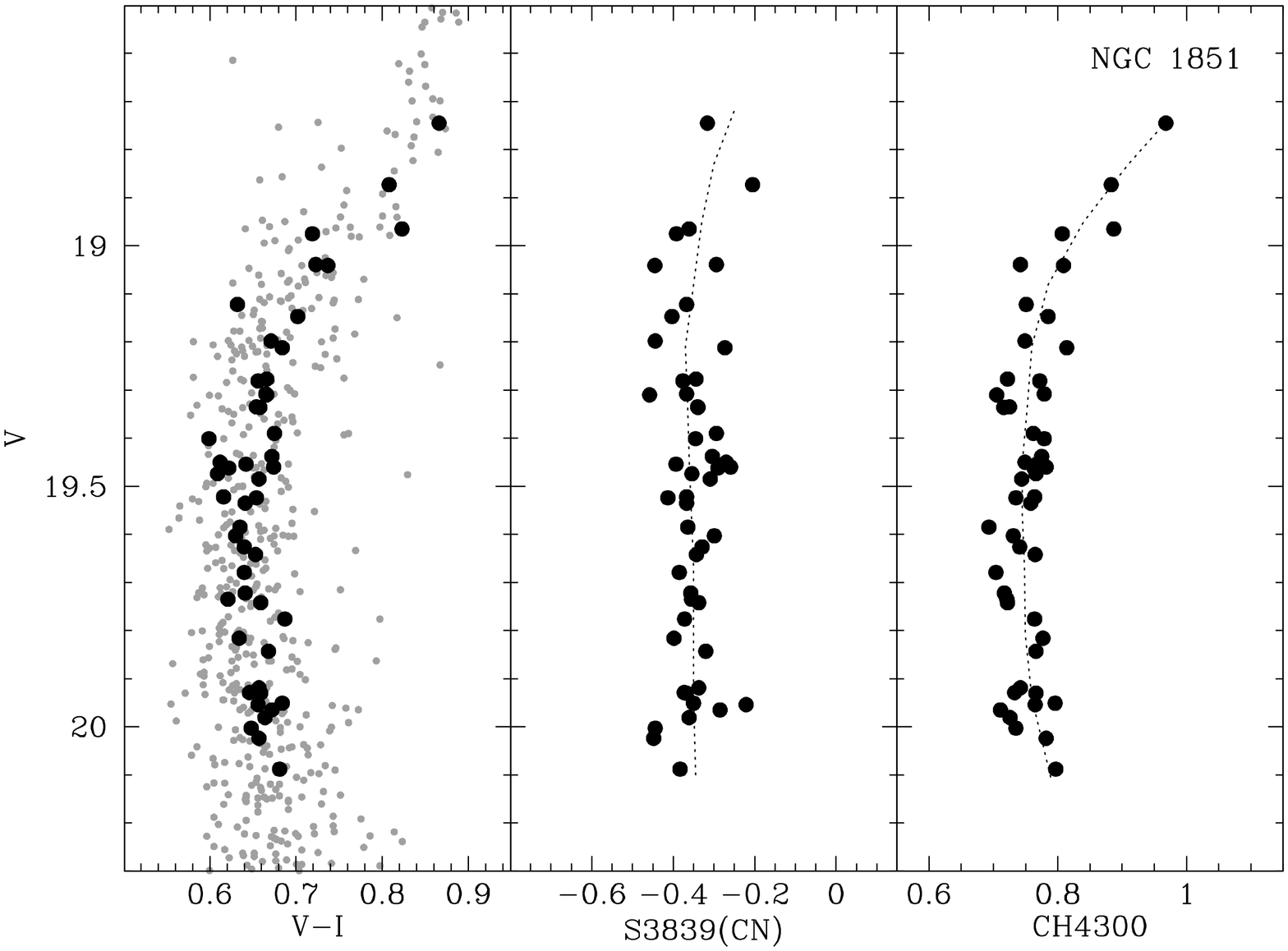}
\caption{Example of the removal of temperature and gravity dependencies
from the CH and CN indices, using median ridge lines on NGC~1851. Large black dots
represent the target stars, while small grey dots represent the calibrated
pre-maging photometry of NGC~1851. The {\em median ridge lines} of CH and CN as
functions of V magnitude are marked as dotted lines.}
\label{fig_tg}
\end{figure}

At a fixed overall abundance, it is well known that the CN-band and CH-band are
stronger in stars with lower temperature and gravity. To eliminate this
dependency, different authors use different proxies for temperature and gravity,
such as colours \citep[e.g.][]{harbeck03} or magnitudes  \citep[e.g.][]{norris81}
or some combination of the Balmer indices \citep[e.g.][]{kayser06}. The indices
are then corrected by fitting the lower envelope of the distribution in the chosen
plane, and the new indices are generally indicated as $\delta$S3839(CN) and
$\delta$CH4300. Most of the cited authors find generally no significant trend for
S3839(CN), but they do always correct the CH4300 index, usually with a
second-order polynomial. Given the diversity of shapes and slopes, we prefer to
use {\em median ridge lines}\footnote{A rough estimate of the uncertainty in the
placement of the median ridge line can be obtained by using the first
interquartile range of the rectified indices, divided by the square root of the
total number of points. We find a typical uncertainty ranging approximately from
0.01 to 0.02 in the CN index, and from 0.005 to 0.01 in the CH index, depending on
the cluster. These uncertainties are negligible for all applications in the
present work.} to correct for the curvature induced by both temperature and
gravity effects. We show an example of the computed median ridge lines, with
NGC~1851 in Fig.~\ref{fig_tg}. 

Generally, we confirm that the curvature of S3839(CN) as a function of V magnitude
is much smaller than the curvature of CH4300. We define our rectified
$\delta$S3839(CN) and $\delta$CH4300 indices as the difference between the
original S3839(CN) and CH4300 indices and their respective median ridge lines. The
rectified indices will be used throughout the rest of this paper.

\begin{figure*}
\centering
\includegraphics[angle=270,width=16cm]{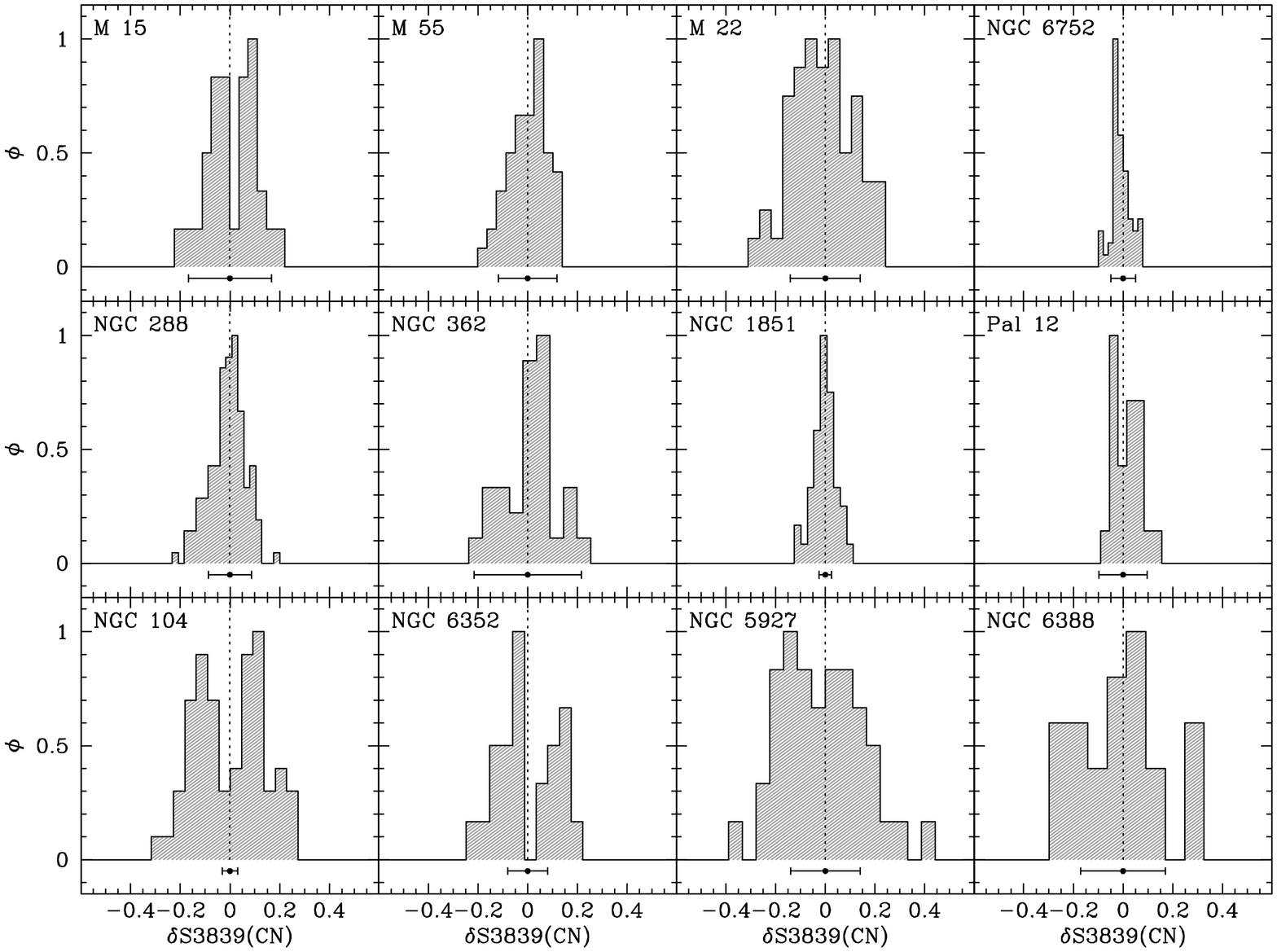}
\includegraphics[angle=270,width=16cm]{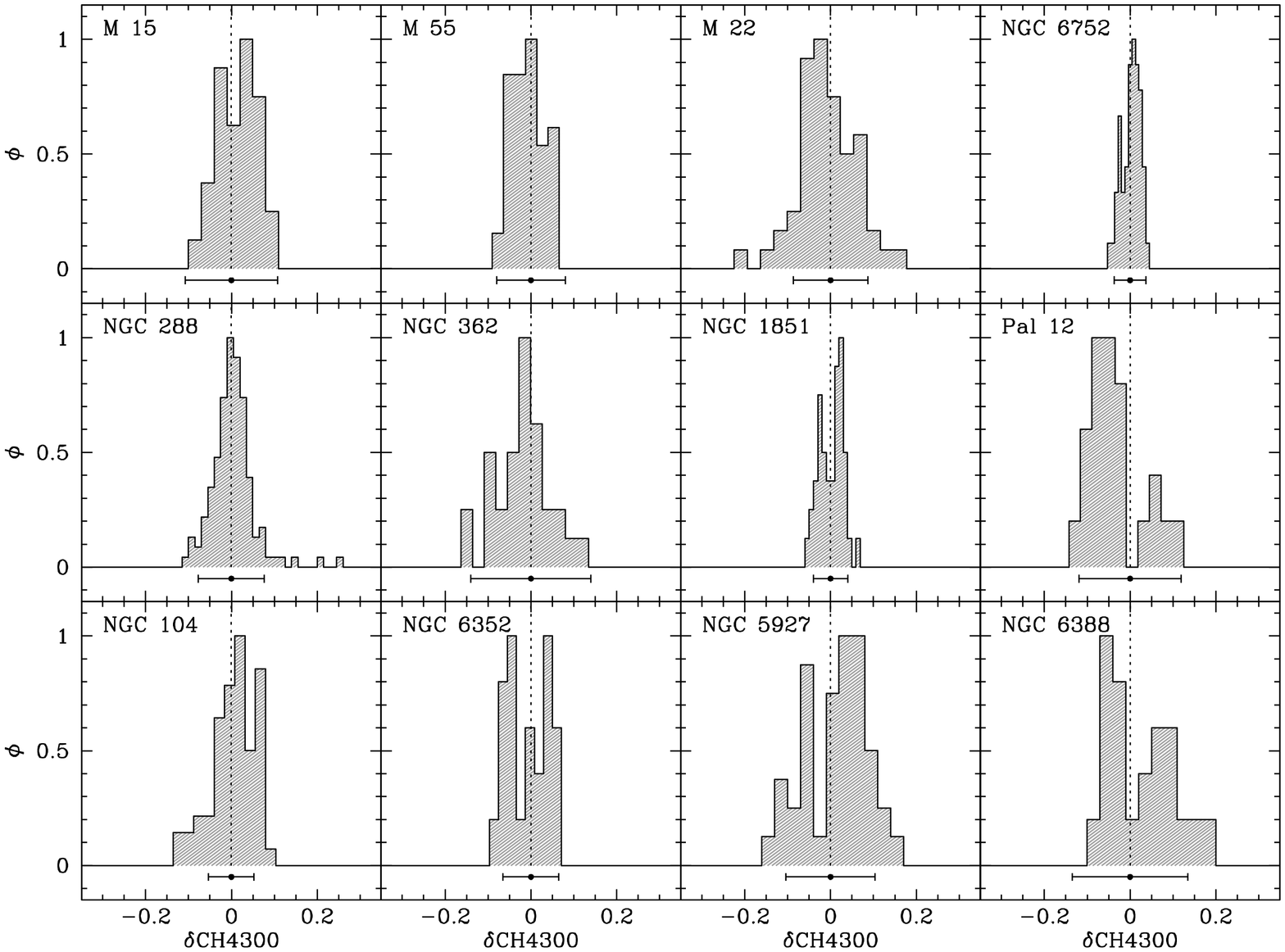}
\vspace{-0.5cm}
\caption{Histograms of the $\delta$S3839(CN) (top panels) and $\delta$CH4300
(bottom panels) index strengths. The bin sizes are larger for clusters with
lower S/N and/or fewer stars. Typical (median) errorbars are plotted below each
histogram. Vertical dotted lines mark $\delta$S3839(CN)=0 and $\delta$CH4300=0,
i.e. the {\em median ridge line} of the rectified index and the border between
CN or CH strong and CN or CH weak stars. Clusters are sorted by increasing
metallicity, from left to right and from top to bottom.}
\label{fig_hist}
\end{figure*}

\begin{figure*}
\centering
\includegraphics[angle=270,width=\textwidth]{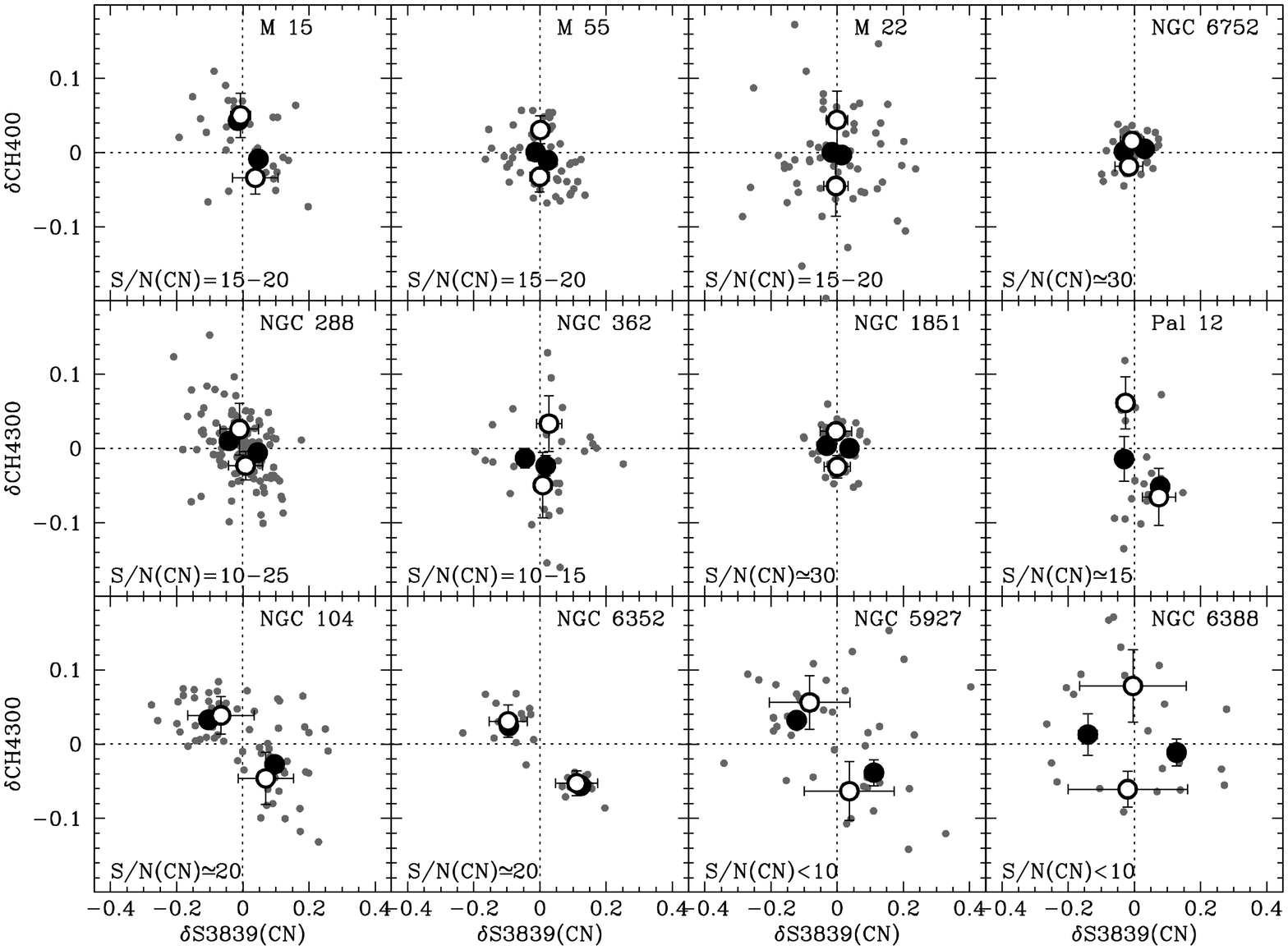}
\vspace{-0.5cm}
\caption{Anti-correlation plots for the CH and CN band strengths. Each panel
shows the measurements for stars (grey dots) in each cluster. CH strong and weak
stars are separted by the horizontal dotted line, and their centroids are marked
as large white dots, while CN strong and weak stars are separated by the vertical
dotted line and their centroids are marked as large black dots. The S/N ratio in
the CN band region is indicated in the lower left corners.}
\label{fig_chcn}
\end{figure*}

\subsection{Additional data from the literature and archives}
\label{sec-kayser}

To increase our observed sample, we re-reduced the 47~Tuc data we found in the ESO
archive with our procedure, which is very similar to the one adopted by the
original team that made the observations \citep{harbeck03}. This way, a direct
comparison of the results was possible, showing excellent agreement and therefore
providing a useful validation of our adopted procedure. We also added index
measurements for the four clusters in \citet{kayser08} that had MS and SGB stars
faint enough to have pristine surface composition, not altered by the first
dredge-up mixing episode, which happens at the base of the RGB. We therefore
selected from their measurements (their Table~A.1) those stars in NGC~362 with
V$<$18.5, in M~55 with V$<$17.5, in M~15 with V$<18.5$, and in M~22 with V$<$17
\citep[see Fig.~2 by][]{kayser08}. Their definition of indices is exactly the same
as ours (Sect.~\ref{sec-index}). We applied a more strict selection in radial
velocity, at 2.5\,$\sigma$ (as in Sect.~\ref{sec-1stqc}). We then removed the
temperature and gravity dependency from their measurements (their Table~A.1),
using the same procedure described in Sect.~\ref{sec-tg}. \citet{kayser08} do not
give the S/N of their spectra, but judging from their errobars (see their
Table~A.1), they should be of the order of S/N$\simeq$15--20 for M~15, M~22, and
M~55, and S/N$\simeq$10--15 for NGC~362 anf NGC~288. In the following sections, we
analyse and discuss these data together with ours.


\section{Results}
\label{sec-res}

\subsection{CH and CN bimodalities}
\label{sec-bimo}

The presence of CH and CN bimodalities, i.e., two well separated groups of stars
with a different strength of these bands, was apparent almost as soon as CH and CN
anomalies were discovered in red giants in the 1970s and 1980s. Some clusters
showed clear bimodalities in CN (or in [N/Fe]) and weaker or missing bimodalities
in CH (or in [C/Fe]), while some other clusters instead showed a continuous spread
in CN strength. Various studies identified CN bimodalities among red giants of M~3
and M~13 \citep{suntzeff81,briley04b}, NGC~6752 \citep{norris81}, $\omega$~Cen
\citep{cohen86}, NGC~6934 \citep{smith86}, NGC~6171 \citep{smith88}, M~71
\citep{smith89,lee05}, M~2 \citep{smith90}, M~5 \citep{ramirez02}, NGC~288 and
NGC~362 \citep{kayser08}, NGC~6121 \citep{marino08}, NGC~6356, and NGC~6528
\citep{martell09}, among others. Bimodalities among MS and sub-giant branch stars
have also been found in M~71 \citep{cohen99}, 47~Tuc
\citep{cannon1998,harbeck03,briley04a,carretta05}, and NGC~6752
\citep{carretta05}.

The interest related to the presence of a bimodal distribution -- rather than a
continuous spread -- is clear in the light of the latest theories of
self-enrichment for GGC (see also Sect.~\ref{sec-intro}). First generation stars
should have a {\em "normal"} surface composition, rich in C and poor in N, while
the second generation ones, formed from CN(O) cycle processing polluted gas,
should have higher N and lower C \citep{smith82}. If we find a clear bimodality in
the data, this supports the idea of two stellar generations\footnote{The opposite
is not necessarily true, in the sense that self-enrichment could also produce a
continuous spread in the CH and CN band strengths, depending on the details of the
chemical enrichment history, such as duration and intensity of star-formation
episodes.}. If a bimodal distribution is found measuring the CH and CN band
strengths,  we expect that the underlying [C/Fe] and [N/Fe] distributions will
turn out to be bimodal as well, since each cluster has the same overall
metallicity and all stars in each cluster have roughly the same atmospheric
parameters. This of course requires confirmation via spectrum synthesis
calculations. These calculations will be the subject of a future paper. Indeed,
such a bimodality in C and N enhancements has been found e.g., by
\citet{briley04b} in RGB stars in M~13, or by \citet{briley04a} in MS stars in
47~Tuc. That such bimodalities have rarely been found among red giants when the
Na-O anti-correlations are considered deserves further studies. We point out two
possible reasons in advance: {\em (i)} the Na and especially the O abundances in
giants are difficult to measure, and indeed we often see only upper limits for
[O/Fe] \citep{car09a,car09b} or {\em (ii)} C, N, O, and Na are not directly
comparable, because they are not produced exactly at the same temperature: C and N
are altered within the CN bi-cycle, O is depleted in the complete CNO cycle and Na
is produced in the NeNa cycle, each dominating at progressively higher
temperatures; in this second hypothesis much care should be taken with the
comparison of Na-O and C-N anti-correlations with each other. Finally we note that
if such a C-N bimodality is present, it should be easier to see in MS stars than
in RGB stars, where it could be attenuated by internal mixing
(Sect.~\ref{sec-intro}). 

\begin{figure}
\centering
\includegraphics[width=\columnwidth]{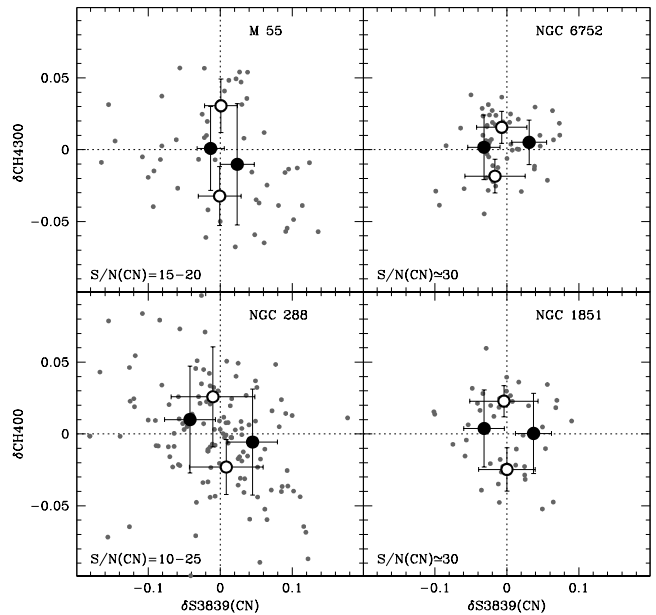}
\caption{Zoomed version of those clusters that show a small spread
on the scale of Fig.~\ref{fig_chcn}.}
\label{fig_zoom}
\end{figure}

We studied the presence of bimodal distributions for the CN and CH rectified band
strengths by means of histograms (Fig.~\ref{fig_hist}) and -- more importantly --
by plotting our measurements in the CH-CN plane (Fig.~\ref{fig_chcn}, see also
next Sect.). We consider a distribution to be ``bimodal" when the centroids of the
CN-strong (CH-weak) and CN-weak (CH-strong) stars are clearly
separated\footnote{This generally implies a significant centroids separation of at
least 1\,$\sigma$, but this separation of centroids does not grant a bimodality
{\em per se}, as for NGC~6388.} in the CH-CN plane and -- as a supporting evidence
-- when the histograms also show signs of bimodality. Detailed studies of the
statistical significance of such bimodalities are deferred to a following paper,
dealing with the [C/Fe] and [N/Fe] distributions. First, we note that all the most
metal-rich clusters (Pal~12, 47~Tuc, NGC~6352, and NGC~5927), with the only
exception of NGC~6388 (which has low S/N and few stars), show a clear bimodality
not only in $\delta$S3839(CN), but also in $\delta$CH4300. Among the metal-poor
clusters, the situation is less clear. There is a bimodality only in M~15, even if
it is the most metal-poor of the sample and has a relatively low S/N ratio. For
the remaining clusters we only see vague hints of possible bimodalities (see
Sect.~\ref{sec-sample} for more details) in Fig.~\ref{fig_hist}, but not in
Fig.~\ref{fig_chcn} or \ref{fig_zoom}. We will discuss individual clusters in
Sect.~\ref{sec-sample}.

\subsection{CN and CH anticorrelations}
\label{sec-anti}

While the S/N of our metal-poor clusters spectra (NGC~362, NGC~6752, NGC~288,
NGC~1851, M~22, M~55) is comparable to that of other authors
\citep[i.e.,][]{harbeck03,kayser08}, the double molecule bands of CN are really
weak at those low metallicities, for MS stars. On the contrary, even if the S/N
of our remaining metal-rich clusters is lower due to bad weather conditions, CH
and CN bands are much stronger. For the MS stars extracted from the
\citet{kayser08} sample, the S/N is relatively low because they focussed their
analysis on brighter RGB and SGB stars. 

Because our S/N is sometimes at the limit of detection in the CN band region, at
least for the most metal-poor clusters, we used the $\delta$CH4300 index, which
has the highest S/N, to separate target stars in a CH-strong ($\delta$CH4300$>$0)
and a CH-weak ($\delta$CH4300$<$0) group for each cluster. A weighted average was
used to compute the centroids of the CH-strong and CH-weak groups in the CN-CH
plane. The resulting centroids with their 1\,$\sigma$ errors are reported in
Fig.~\ref{fig_chcn} as large white dots along with measurements for each star. For
completeness we also divided the stars in CN-strong ($\delta$S3839(CN)$>$0) and
CN-weak ($\delta$S3839(CN)$<$0) groups, and their centroids are marked as large
black dots with their errobars. A zoom-in for those clusters with small spreads on
the scale of Fig.~\ref{fig_chcn} is presented in Fig.~\ref{fig_zoom}.

We see clear anti-correlations (Fig.~\ref{fig_chcn}) only for those clusters
that showed a clear bimodality in both indices, i.e., M~15, Pal~12, 47~Tuc,
NGC~6352, and NGC~5927. We would be tempted to say that where our data are good
enough to show an anticorrelation, this anticorrelation is always clearly
bimodal. The case of NGC~288 seems to be an exception, and it is further
discussed in Sect.~\ref{sec-sample}. In most of the other cases,
anti-correlations and/or bimodalities were found by other authors, as discussed
in detail in Sect.~\ref{sec-sample}, so we suspect that in general the
present data do not have high enough S/N to be conclusive on these clusters.

\begin{table*}
\caption{Clusters sample intrinsic properties and structural parameters.}
\label{int_params}
\begin{center}
\begin{tabular}{l c c c c c c c c c c c c c c c c c c} 
\hline \hline
Name & [Fe/H] & HBR &
log(T$_{HB}^{max}$) & Age$^{\mathrm{l}}$ &
$\varepsilon$ & $\sigma_{\mathrm{o}}$ &  c & $r_t$ & $r_h/r_J^{\mathrm{p}}$ &
M$_V^{Tot}$ & log($\frac{M}{M_{\odot}})$ & M/L \\ 
& (dex) && ($^{\mathrm{o}}$K) & (Gyr) && (km\,s$^{-1}$) & ($^{\prime}$) & ($^{\prime}$) 
&& (mag) & (dex) & (M$_{\odot}$\,L$_{\odot}^{-1}$) \\
\hline
NGC 104 (47~Tuc) & -0.75$^{\mathrm{a}}$ & -0.99 	       & 3.756$^{\mathrm{i}}$ & 13.06 & 0.09		    & 11.5$^{\mathrm{n}}$ & 2.03 & 42.86 & ---   & -9.42 & 6.05$^{\mathrm{q}}$ & 1.33$^{\mathrm{q}}$ \\
NGC 288          & -1.26$^{\mathrm{a}}$ &  0.98 	       & 4.221$^{\mathrm{j}}$ & 10.62 & 0.09$^{\mathrm{m}}$ &  2.9$^{\mathrm{n}}$ & 0.96 & 12.94 & 0.093 & -6.74 & 4.85$^{\mathrm{q}}$ & 2.15$^{\mathrm{q}}$ \\
NGC 362          & -1.33$^{\mathrm{b}}$ & -0.87$^{\mathrm{h}}$ & 4.079$^{\mathrm{i}}$ & 10.37 & 0.01		    &  6.4$^{\mathrm{n}}$ & 1.94 & 16.11 & 0.023 & -8.41 & 5.53$^{\mathrm{q}}$ & 0.90$^{\mathrm{r}}$ \\
NGC 1851         & -1.22$^{\mathrm{c}}$ & -0.32$^{\mathrm{h}}$ & 4.097$^{\mathrm{i}}$ &  9.98 & 0.05		    & 10.4$^{\mathrm{n}}$ & 2.32 & 11.70 & 0.015 & -8.33 & 5.49$^{\mathrm{q}}$ & 1.61$^{\mathrm{q}}$ \\
NGC 5927         & -0.67$^{\mathrm{d}}$ & -1.00 	       & 3.724$^{\mathrm{i}}$ & 12.67 & 0.04		    &  4.3$^{\mathrm{o}}$ & 1.60 & 16.68 & ---   & -7.80 & 5.32$^{\mathrm{r}}$ & ---  \\
NGC 6352         & -0.55$^{\mathrm{e}}$ & -1.00 	       & ---		      & 12.67 & 0.07		    &  5.4$^{\mathrm{o}}$ & 1.10 & 10.51 & ---   & -6.48 & 4.57$^{\mathrm{r}}$ & ---  \\
NGC 6388         & -0.41$^{\mathrm{a}}$ & -0.69$^{\mathrm{h}}$ & 4.255$^{\mathrm{i}}$ & 12.03 & 0.01		    & 18.9$^{\mathrm{n}}$ & 1.70 &  6.21 & ---   & -9.42 & 6.02$^{\mathrm{q}}$ & 1.89$^{\mathrm{q}}$ \\
NGC 6656 (M~22)  & -1.76$^{\mathrm{f}}$ &  0.91$^{\mathrm{h}}$ & ---		      & 12.67 & 0.14		    &  9.0$^{\mathrm{n}}$ & 1.31 & 28.97 & ---   & -8.50 & 5.56$^{\mathrm{q}}$ & 2.07$^{\mathrm{q}}$ \\
NGC 6752         & -1.56$^{\mathrm{a}}$ &  1.00 	       & 4.471$^{\mathrm{j}}$ & 11.78 & 0.04		    &  4.5$^{\mathrm{n}}$ & 2.50 & 55.34 & ---   & -7.73 & 5.16$^{\mathrm{r}}$ & ---  \\
NGC 6809 (M~55)  & -1.97$^{\mathrm{a}}$ &  0.87 	       & 4.153$^{\mathrm{j}}$ & 12.29 & 0.02		    &  4.9$^{\mathrm{n}}$ & 0.76 & 16.28 & ---   & -7.55 & 4.99$^{\mathrm{q}}$ & 3.23$^{\mathrm{q}}$ \\
NGC 7078 (M~15)  & -2.34$^{\mathrm{a}}$ &  0.67$^{\mathrm{h}}$ & 4.477$^{\mathrm{i}}$ & 12.93 & 0.05		    & 12.0$^{\mathrm{n}}$ & 2.50 & 21.50 & 0.027 & -9.17 & 5.84$^{\mathrm{r}}$ & ---  \\
Pal 12           & -0.80$^{\mathrm{g}}$ & -1.00 	       & ---		      &  8.83 & ---		    &  1.3$^{\mathrm{o}}$ & 1.94 & 17.42 & 0.193 & -4.48 & 3.75$^{\mathrm{q}}$ & ---  \\
\hline\hline
\end{tabular}
\end{center}
{\bf NOTES:} All parameters are derived from the 2003 revision of the
\citet{harris96} GGC catalogue except: 
$^{\mathrm{a}}$ \citet{car09c};  
$^{\mathrm{b}}$ \citet{shetrone00};
$^{\mathrm{c}}$ \citet{yong09};
$^{\mathrm{d}}$ \citet{kraft03};
$^{\mathrm{e}}$ \citet{feltzing09};
$^{\mathrm{f}}$ \citet{marino09};
$^{\mathrm{g}}$ \citet{cohen04};
$^{\mathrm{h}}$ \citet{catelan09};
$^{\mathrm{i}}$ \citet{recioblanco06};
$^{\mathrm{j}}$ Bragaglia (2009, private communication);
$^{\mathrm{l}}$ relative ages by \citet{marin-franch09} converted to absolute 
ages multiplying by 12.8~Gyr;
$^{\mathrm{m}}$ \citet{frenk82};  
$^{\mathrm{n}}$ \citet{pryor93};
$^{\mathrm{o}}$ \citet{gnedin02};  
$^{\mathrm{p}}$ \citet{baum09}; 
$^{\mathrm{q}}$ \citet{mclaughlin05};
$^{\mathrm{r}}$ \citet{mandushev91}.\\
{\bf Meaning of columns.} (1) cluster name (alternate name in parenthesis); (2) mean iron
abundance ratio; (3) HB morphology, where HBR=(B-R)/(B+V+R), being B the number of stars
bluer than the instability strip, R redder, and V the number of variables in the strip; (4)
the maximum temperature of the HB; (5) the age in Gyr; (6) the isophotal ellipticity
$\epsilon$=1-(b/a); (7) the central radial velocity dispersion; (8) the concentration (2.5
means core collapsed cluster); (9) the tidal radius; (10) the ratio between the half-light
and the Jacobi radius; (11) the integrated V magnitude; (12) the logarithm of the total
mass in solar units; (12) the mass-to-light radius in solar units.  
\end{table*}

\section{Cluster by cluster discussion}
\label{sec-sample}

\subsection{NCG~104 (47~Tucanae)}

The cluster NGC~104 or 47~Tuc is one of the most massive and metal-rich GGC
(Table~\ref{int_params}). It is considered to be part of the thick disc
\citep{dinescu+99}, with a low eccentricity and relatively high inclination orbit.
Currently it is located $\simeq 3$~kpc below the Galactic plane in projection  on
the sky close to the Small Magellanic Cloud.  \citet{pritzl05} place it in the
group of thick-disc clusters, while \citet{mackey+gilmore04} classify it as
bulge/disc cluster. 

It is one of the most studied GGC thanks to its vicinity and its relatively high
[Fe/H]. While the abundance determinations on red giants are too many to mention
here \citep[the most recent being][]{car09c}, the abundance analysis of turn-off
and early subgiant stars resulted in [Fe/H]=--0.67 $\pm$0.01($\pm$0.04)~dex
\citep{carretta+04}, while \citet{james+04} derive --0.69 $\pm$0.06~dex for SGB
stars and --0.68$\pm$0.01~dex from three turnoff stars.  It was one of the first
GGC in which CN-CH variations were detected among MS stars
\citep{cannon1998,briley94}. The star-to-star scatter has been clearly detected
for light elements; while the Na-O anticorrelation is not very strong, it has
nevertheless been detected by \citet{carretta+04}. 

The CN-CH molecular abundances of turn-off and sub-giant stars in this cluster
were also topic of a detailed study by \citet{harbeck03} and \citet{briley04a},
resulting in a clear bimodality in band strengths and also in [N/Fe],  with
abundance differences of up to 1~dex. We use here their archive data to test the
homogeneity of reductions. A clearly bimodal anti-correlation is detected here
(Fig.~\ref{fig_chcn}), which confirms the results by \citet{harbeck03}  and
\citet{briley04a} and which, together with the double SGB and wide MS found by
\citet{anderson09}, demonstrates the presence of two distinct populations in this
cluster.

\subsection{NGC~288}

The source NGC 288 is a low concentration, low central density cluster located
close to the South Galactic Pole, in a very low interstellar extinction region.
Its retrograde motion and relatively low velocity components place it among the
inner old halo (OH) globular clusters \citep{dinescu+97}. Tidal shocks due to
bulge and disc for this cluster are estimated to be strong compared to the usually
more predominant internal relaxation and evaporation \citep{dinescu+99}.

Its horizontal branch is composed almost entirely of stars bluer than the 
instability strip. It is often studied together with NGC 362, which has the same
metallicity, but very different HB morphology: they represent one of the classical
second parameter pairs. \citet{bolte89} was one of the first to propose helium,
CNO, or age differences as possible explanations for the second parameter problem.
\citet{shetrone00} made a comparative spectroscopic study of red giants in the
second-parameter GC pair NGC 288 and NGC 362. The [Fe/H] they derived for NGC~288
is $-1.39 \pm 0.01$~dex. These authors also remark on the Na-O and Al-O
anti-correlations, which were found in both clusters. Fig.~7 by \citet{car09b}
shows even a hint of bimodality in the Na-O anti-correlation. A clear bimodality
in the CH and CN index strength has been detected among red giants by
\citet{kayser08} and  \citet{smith09}.

Our data are the ones by \citet{kayser08}, of which we selected only unevolved
stars, and we do not see any clear anti-correlation. There is no reason why
unevolved stars in a GGC should have a uniform composition while the evolved ones
show a bimodality, because no known mixing mechanism would be able to create a
bimodality from a homogeneous population. Therefore we must conclude that our data
do not have a sufficient S/N to reveal a bimodality in these metal-poor MS stars.
Spectral synthesis calculation will be able to quantify the sensitivity of our
spectra to C and N abundances, while more higher quality data are most probably
needed anyway to settle the question. We note here that as we will see in
Sect.~\ref{sec-trends}, NGC~288 falls perfectly in line with other clusters in all
correlations with cluster parameters.

\subsection{NGC~362}

As mentioned before, NGC~288 and NGC~362 are one of the second-parameters pairs,
with similar metallicity and different HB morphology. \citet{bolte89} proposed
CNO, among others, as one of the second parameters. NGC~362 has a metallicity of
--1.33$\pm$0.01~dex \citep{shetrone00}, it is more concentrated than NGC~288
(Table~\ref{int_params}) and slightly closer to the Galactic centre. The anomalies
in the CH and CN band strengths among its giants were studied more than in NGC~288
\citep{mcclure74,frogel83,smith83,smith84}. \citet{kayser08} extensively studied
both NGC~288 and NGC~362, finding a clear anti-correlation among RGB and SGB stars
in both clusters, and a clear bimodality among red giants {\em in both clusters}.
This suggests that anti-correlations or chemical anomalies cannot be the
predominant source of the different HB morphology in these two clusters. 

As for NGC~288, bimodalities and an anti-correlation were found in the CH and CN
band strengths of RGB stars, while our data for the MS stars reveal none. We
suspect that here also this is simply owing to the low S/N of the spectra (lower
than for NGC~288), and it must be worsened by the smaller sample. A study with
higher quality spectra would be even more interesting than for NGC~288, because
NGC~362 does not behave like the other GGC in the correlations with cluster
parameters (Sect.~\ref{sec-trends}) and it is not clear whether NGC~362 is a real
outlier or if the present data are simply inconclusive.

\subsection{NGC~1851}
\label{sec-1851}

The cluster NGC~1851 has a very eccentric and highly inclined orbit, with a large
excursion into the outer parts of the Galaxy \citep{dinescu+97,dinescu+99}. The
iron abundance measurements from high-resolution spectra report [Fe/H]=$-1.22 \pm
0.03$~dex \citep{yong09} and $-1.27 \pm 0.09$ \citep{yong08}. The bimodal HB in
this cluster already hinted towards multiple stellar populations, but only
recently \citet{milone08} detected two parallel SGB sequences, thanks to exquisite
ACS@HST photometry, and \citet{han09} found that the use of U, U--I photometry
reveals a striking bimodality in the RGB. A possible different radial distribution
of the two SGB \citep{zoccali+09} would support some predictions of the {\em
self-enrichment scenario} \citep{dercole08}, but \citet{milone09} do not confirm
this finding.

A possible explanation for the double SGB is a CNO enriched stellar component
\citep{cassisi08, salaris08}, possibly -- but not necessarily -- combined with
some age difference of the order of 1~Gyr at most. This is expected to produce a
strong effect on CN and  CH molecules. Indeed, among the RGB stars some extremely
CN strong stars have been detected \citep{hesser+82}.  Another study reporting a
large CNO abundance spread in giant stars is that by \citet{yong09}. The HB
morphology of NGC~1851 seems to rule out a large He difference between the two
populations \citep{salaris08}.

Unfortunately, as for NGC~288 and NGC~362, at this metallicity our S/N ratio is
probably not sufficient to reveal anti-correlations and bimodalities. We only see
some bimodality in the $\delta$CH4300 index (the one with the highest S/N), but
not in $\delta$S3839(CN), and only in Fig.~\ref{fig_hist}, but not in
Fig.~\ref{fig_chcn} or \ref{fig_zoom}. Moreover, our targets do not fall in the
ACS field, so we cannot attempt any experiment on their position on the two
different SGB found by \citet{milone08} at this stage. More studies including
higher S/N spectra and a careful target selection could help in clarifying the
nature of these substructures. As for NGC~288, NGC~1851 behaves like all other
clusters of similar metallicity when trends with GGC parameters are considered
(Sect.~\ref{sec-trends}).

\begin{table}
\caption{Clusters sample positional and orbital parameters.}
\label{ext_params}
\begin{center}
\setlength{\tabcolsep}{1.5mm}
\begin{tabular}{l c c c c c c c c c c c c c c c c c c} 
\hline \hline
Name & Class$^{\mathrm{a,b}}$  & R$_{GC}$ & E$_{tot}$ & z$_{max}$ &  $\Psi$ & P
\\  
& & (kpc) & (10$^2$km$^2$s$^{-2}$) & (kpc) & (deg) & (10$^6$yr) \\
\hline
NGC 104  & BD (TK)  &  7.4$^{\mathrm{c}}$ &  -872$^{\mathrm{e}}$ & 3.1$^{\mathrm{e}}$ & 29$^{\mathrm{e}}$ & 190$^{\mathrm{e}}$ \\
NGC 288  & OH (H)   & 12.0$^{\mathrm{c}}$ &  -787$^{\mathrm{e}}$ & 5.8$^{\mathrm{e}}$ & 44$^{\mathrm{e}}$ & 224$^{\mathrm{e}}$ \\
NGC 362  & YH (H)   &  9.4$^{\mathrm{c}}$ &  -856$^{\mathrm{e}}$ & 2.1$^{\mathrm{e}}$ & 21$^{\mathrm{e}}$ & 208$^{\mathrm{e}}$ \\
NGC 1851 & OH (---) & 16.7$^{\mathrm{c}}$ &  -340$^{\mathrm{e}}$ & 7.6$^{\mathrm{e}}$ & 22$^{\mathrm{e}}$ & 584$^{\mathrm{e}}$ \\
NGC 5927 & BD (---) &  4.5$^{\mathrm{c}}$ & -1020$^{\mathrm{f}}$ & 0.7$^{\mathrm{f}}$ &  9$^{\mathrm{f}}$ & 147$^{\mathrm{f}}$ \\
NGC 6352 & BD (TN)  &  3.3$^{\mathrm{c}}$ &  ---  & --- & ---& --- \\
NGC 6388 & BD (---) &  3.2$^{\mathrm{c}}$ &  ---  & --- & ---& --- \\
NGC 6656 & OH (TK)  &  4.9$^{\mathrm{c}}$ &  -871$^{\mathrm{e}}$ & 1.9$^{\mathrm{e}}$ & 18$^{\mathrm{e}}$ & 190$^{\mathrm{e}}$ \\
NGC 6752 & OH (TK)  &  5.2$^{\mathrm{c}}$ &  -977$^{\mathrm{e}}$ & 1.6$^{\mathrm{e}}$ & 18$^{\mathrm{e}}$ & 156$^{\mathrm{e}}$ \\
NGC 6809 & OH (H)   &  3.9$^{\mathrm{c}}$ & -1038$^{\mathrm{e}}$ & 3.7$^{\mathrm{e}}$ & 56$^{\mathrm{e}}$ & 122$^{\mathrm{e}}$ \\
NGC 7078 & YH (H)   & 10.4$^{\mathrm{c}}$ &  -752$^{\mathrm{e}}$ & 4.9$^{\mathrm{e}}$ & 36$^{\mathrm{e}}$ & 242$^{\mathrm{e}}$ \\
Pal 12   & SG (H)   & 16.2$^{\mathrm{d}}$ &   650$^{\mathrm{d}}$ &20.1$^{\mathrm{d}}$ & 58$^{\mathrm{d}}$ & 730$^{\mathrm{d}}$ \\
\hline\hline
\end{tabular}
\end{center}
{\bf NOTES:} Literature references are: 
$^{\mathrm{a}}$ GGC classification by \citet{mackey05}, where BD=bulge/disk, 
OH=old halo, YH=young halo, SG=Sagittarius;
$^{\mathrm{b}}$ GGC classification (in paranthesis) by \citet{pritzl05}, where 
TK=thick disk, TN=thin disk, H=halo;
$^{\mathrm{c}}$ 2003 revision of the \citet{harris96} catalogue;
$^{\mathrm{d}}$ \citet{dinescu+00};
$^{\mathrm{e}}$ results with model JSH95 by \citet{dinescu+99};
$^{\mathrm{f}}$ \citet{dinescu07}\\
{\bf Meaning of columns.} (1) cluster name; (2) classification according to Galactic
sub-component; (3) Galactocentric radius; (4) total orbit energy; (5) maximum height above
the Galactic plane; (6) orbit inclination; (7) orbital period.
\end{table}

\subsection{NGC~5927}
\label{sec-5927}

The source NGC~5927 belongs to the metal-rich globular clusters. It suffers from
relatively high reddening and thus was not well studied spectroscopically with
high-resolution data: according to \citet{zw84}, it is one of the most metal-rich
disc clusters ([Fe/H]=$-0.3$~dex); \citet{cohen83} found this cluster to be
+0.59~dex more metal rich than 47 Tuc; \citet{kraft03} derive from low-resolution
spectra [Fe/H]$_{II}$=$-0.67$~dex; \citet{francois91} derived [Fe/H] $-1.08$~dex
based on the high-resolution study of one star only. Given the large spread in
literature determinations, we adopt the \citet{kraft03} estimate, keeping in mind
that NGC~5927 could well be more metal-rich than that. To the best of our
knowledge, the present study is the first one dedicated to light element
anti-correlations in this cluster. 

We find a quite clear and bimodal anti-correlation of the CH and CN band strength,
in spite of the low S/N ratio of our spectra and of the relatively high reddening
in the cluster field. Clearly, at this metallicity the CN and CH bands are so
strong that it is quite easy to reveal these variations. Finding a bimodal
anti-correlation in such a metal-rich cluster is interesting, since for example no
open cluster revealed these variations up to now \citep[see][and references
therein]{desilva09,panci09,martell09}. Obviously, high-resolution spectroscopy of
stars in this neglected cluster would be of enormous interest.

\subsection{NGC~6352}
\label{sec-6352}

NGC 6352 is a sparsely populated metal-rich cluster with disc kinematics. 
\citet{pritzl05} place it in the group of thin disc clusters, while
\citet{mackey+gilmore04} classify it as a bulge/disc GGC. Based on high-resolution
spectra of eight stars, \citet{cohen83} found this cluster to be +0.38~dex more
metal rich than 47~Tuc. \citet{carretta+gratton97} re-analysed equivalent width
measurements of \citet{gratton87} for three stars finding an average
[Fe/H]$=-0.64\pm 0.06$, and recently \citet{car09c} re-adjusted it to
[Fe/H]$=-0.62\pm 0.05$. A recent high-resolution study by \citet{feltzing09},
based on nine horizontal branch stars, found [Fe/H]$=-0.55 \pm 0.03$ and suggested
that on the HB there is a correlated trend in Al and Na abundances, similarly to
what was found for  stars on the RGB in other globular clusters.

No study dedicated to light elements was carried out to our knowledge, so the hint
by \citet{feltzing09} is our only comparison. We do find a clear and bimodal CH
and CN anti-correlation, which is even more striking if one considers the low
number of stars observed and the relatively low S/N around 3900~\AA. As for
NGC~5927, the high metallicity helps in revealing CH and CN band strength
variations, and the study of these metal-rich GGC would be interesting for the
same reasons. An HST photometric study -- seeking for multiple sequences --
coupled with high-resolution spectroscopy for deriving the chemical patterns of
NGC~6352 could be one of the next steps in the study of chemical abundance
anomalies in GGC.

\subsection{NGC~6388}

The cluster NGC 6388 is one of 10 most massive in the Milky Way, it is a bulge
cluster, located about 3~kpc from the Galactic centre. It is very centrally
concentrated and tightly bound, and has among the highest predicted escape
velocity at the cluster  center \citep{mclaughlin05}. Based on high-resolution
FLAMES-UVES spectra, \citet{carretta07b} measured [Fe/H]$=-0.44 \pm 0.01 (\pm
0.03)$. These authors also detected the presence of Na-O and Mg-Al
anti-correlations, and of a Na-Al correlation among RGB stars. A clear bimodality
in CH and CN was also detected on the RGB by \citet{smith09}. 

NGC~6388 is sometimes referred to as being unusual \citep[e.g.][]{dalessandro08}:
in contrast to expectations for its high metallicity, the cluster harbours  an
extended blue horizontal branch \citep{rich93}, and RR Lyr stars with periods much
longer than expected for their metallicity. In addition, the horizontal branch
presents a slope, so that in the V-band its blue tail lies about 0.5 mag brighter
than the red HB clump \citep{raimondo02}. These features are not reproducible by
stellar evolutionary models with an SSP of standard GGC abundance ratios, but
could be explained with the presence of more than one stellar population and some
self-enrichment with CNO processed material \citep{yoon08}. The evidence for
multiple stellar populations came from the most recent deep near-IR photometry of
\citet{moretti09}, who detected two distinct sub-giant branches in this cluster.

Unfortunately, the strong differential reddening and the insufficient S/N of our
data prevent us from reaching any conclusion about the CH and CN correlation at
the MS level. But given its high metallicity and total mass, this is one of the
most interesting clusters for further investigations. In spite of the poor data
for this cluster, NGC~6388 follows all the trends of other GGC with clusters
parameters (Sect.~\ref{sec-trends}).

\subsection{M~22 (NGC~6656)}
\label{sec-m22}

Besides $\omega$~Centauri, M~22 is one of the first GGC that were suspected to
host multiple stellar populations and chemical anomalies \citep{hesser77}.
Unfortunately, strong differential reddening has complicated its study
\citep{richter99}, therefore spectroscopic studies initially gave conflicting
results:  while a spread in the CH and CN abundances of RGB stars appeared
unquestionable \citep{norris83}, some studies reported on abundance variations of
0.3$\pm$0.5~dex in Ca and/or Fe, often correlated with the CH and CN variations
\citep{peterson80,pilachowski82,lehnert91,brown92}, while other studies found no
significant variation in the heavy element content
\citep{manduca78,cohen81,gratton82,laird91}. This was probably because the
reported variations were of the order of the quoted uncertainties \citep[see
also][]{monaco04}. Recently, \citet{marino09} found that two groups of stars exist
in M~22, where [Fe/H] appears correlated with the $s$-process elements and with
[Ca/Fe], and the usual Na, O, Al anti-correlations were found. \citet{dacosta09}
confirm that the Ca distribution shows a large spread with a possible bimodality. 

Our data, belonging to the unevolved stars from the sample by \citet{kayser08} do
not have sufficient S/N to reveal CH and CN anti-correlations or bimodalities. No
clear bimodality was found by \citet{kayser08} among RGB stars, either. If really
different metallicities co-exist in the cluster, this could further complicate the
detection of anti-correlations and of their bimodalities, if they are present.
Most probably spectral synthesis and the determination of [C/Fe] and [N/Fe] would
help in clarifying the picture.

\subsection{NGC~6752}
\label{sec-6752}

The cluster NGC~6752 is one of the best studied regarding its chemical properties,
given its relative proximity to us (Table~\ref{logs}). It has an intermediate
metallicity and a blue HB. Its orbit is remarkably similar to that of 47~Tuc
\citep{dinescu+00}, and it is therefore associated with the thick disc by some
authors. Many high-resolution spectroscopic studies of this cluster provide the
following iron abundance measurements:  [Fe/H]$=-1.48 \pm 0.01 \pm 0.06$ dex by
\citep{gratton05} based on FLAMES-UVES spectra of 7 giants near the RGB bump,
[Fe/H]$=-1.42$  based on UVES spectra of MS and sub-giant stars \citep{gratton01},
[Fe/H]$=-1.49 \pm 0.07$ for 9 SGB stars, and [Fe/H]$=-1.48 \pm 0.07$ for 9 turnoff
stars based on FLAMES-UVES spectra \citep{james+04},  [Fe/H]$=-1.61$ based on 38
RGB stars observed at very high resolution \citep{yong05}, and  [Fe/H]$_I=-1.56$,
[Fe/H]$_{II}=-1.48$  based on FLAMES GIRAFFE  spectra of  137 RGB stars
\citep{carretta07,car09c}. 

The O-Na anticorrelation is well established in NGC~6752 from observations of
both MS and SGB \citep{gratton01, carretta05}, as well as of RGB
\citep{norris95,yong03}. Data from \citet{grundahl02} and \citet{yong03} show 
Na-O and Al-Mg anticorrelations in stars that are both brighter and fainter than
the RGB bump. \citet{pasquini08} detected very strong N abundance variations in
the MS Turn-off stars. Furhtermore, \citet{yong08} detected for the first time
correlations of light elements with Si and heavier elements among red giants.
Recently, \citet{villanova09} have obtained a direct He measurement from the
5875~\AA\  line in HB stars and found a homogeneous abundance of
Y=0.245$\pm$0.012, thus putting in doubt He as the origin of multiple
populations, at least in this GGC.

Unfortunately, our data have certainly too low a S/N to clearly detect
anti-correlations. The histograms in Fig.~\ref{fig_hist} are suggestive of a
possible bimodality, but we know already that the anti-correlations are bimodal in
this cluster because \citet{norris81} found a clear bimodality among RGB stars and
\citet{carretta05} among MS stars.

\subsection{M~55 (NGC~6809)}

The source M~55 is a moderately sized cluster (Table~\ref{int_params}), with an
unusually low central concentration, which makes it relatively easy to collect
stellar samples even within the cluster core. The cluster lies in front of the 
southern tail of the Sagittarius dwarf galaxy \citep{lane09}. There are few
spectroscopic measurements of metallicity for M~55 stars, yielding [Fe/H]$=$--1.65
\citep{caldwell88},  [Fe/H]$=$--1.95 \citep{minniti93}, and recently
[Fe/H]$=$--1.93$\pm$0.02($\pm$0.07)~dex from 14 stars observed with UVES@VLT
\citep{car09c}, who also found a Na-O anti-correlation and some indication of
possible Al variations. \citet{marin-franch09} assign M~55 to the old group of
GGC, based on their relative ages scale. The blue straggler stars (BSS) of M~55
suggest an unusually high primordial population of binaries compared to other GGC
\citep{lanzoni07}, although this contrasts with the conclusions of
\citet{sollima07}. Another interesting result concerns the abundance of helium,
measured with the R-method by \citet{vargas07}, which appears to be the highest
recorded for a massive GGC: Y=0.274$\pm$0.016.

Our results on this cluster are somewhat unclear, with a possible secondary peak
in the CH band histograms, but not in the CN ones (Fig.~\ref{fig_hist}). This is
most probably owing to the insufficient S/N ratio for MS stars in this metal-poor
cluster. However, \citet{kayser08} found no clear sign of anti-correlation, even
considering the higher S/N red giants spectra, and an early paper by
\citet{briley93} suggests that the CN spread might be unusually low in this
cluster. So the question of CH and CN anti-correlations and of their bimodalitiy
in M~55 remains open.

\begin{figure}
\centering
\includegraphics[angle=270,width=\columnwidth]{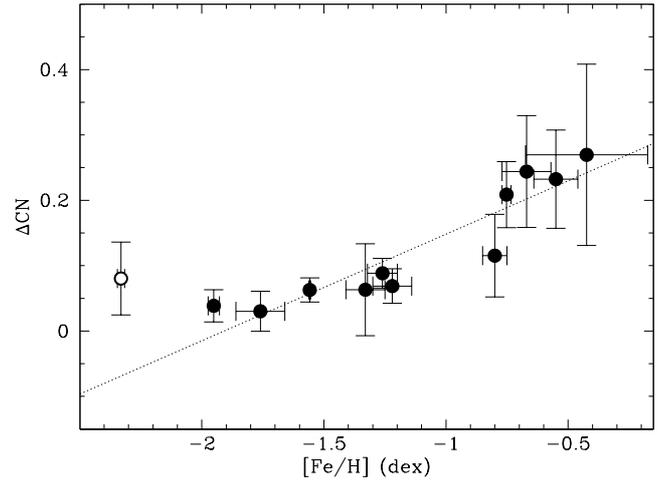}
\caption{Relation between the CN-strong and CN-weak centroids geometrical
distance in the CN-CH plane and [Fe/H] from Table~\ref{int_params}. The white dot
represents M~15, which is excluded from the linear fit (dotted line) of the
remaining clusters (black dots).} 
\label{fig_dist}
\end{figure}

\subsection{M~15 (NGC~7078)}
\label{sec-m15}

The most metal-poor cluster in our sample is M~15, with [Fe/H]=--2.34
\citep{car09c}, it has a blue HB and it is a core-collapsed GGC, which is
suspected to harbour a black hole in its centre \citep[see][and references
therein]{ho03,macnamara03,van06,chakra06,kiselev08}. It is at present quite far
away from the Galactic center (Table~\ref{ext_params}).

Anti-correlations were detected among all usual light elements in M~15
\citep[see][and references therein]{sneden97}, with hints that there could be a
spread also in the heavy elements \citep[see also][]{sneden00,otsuki09}. On the
low-resolution spectroscopy front, \citet{cohen05} obtained a beautiful C-N
anti-correlation among stars on the lower RGB, with impressive abundance
variations of 1--2~dex. However, neither their [C/Fe] nor [N/Fe]
distributions show any evidence of bimodality. Also, \citet{lee00} found an
anti-correlation among bright giants. The Na-O anti-correlation was also found
by  \citet{car09b}, although their measurements were mostly upper limits.

We selected our targets among the unevolved stars measured by \citet{kayser08},
who found some bimodality among both evolved and unevolved stars. With our median
ridge line correction of the temperature and gravity effect (Sect.~\ref{sec-tg})
we were able to reveal a clearly bimodal CH and CN anti-correlation. This is
really striking if one considers the very low metallicity and the S/N ratio of
these spectra, lower than that of NGC~1851 and comparable to that of NGC~288.
Indeed, the index strengths measured in this paper are -- as expected -- a
function of global metallicity, represented by [Fe/H] in
Fig.~\ref{fig_dist}\footnote{$\Delta$CN is the geometrical centroids distance of
the CN-strong and CN-weak groups in the CH-CN plane. The geometrical distance is
$d=\sqrt(x^2+y^2)$, where $x$ is the difference between the average
$\delta$S3839(CN) of the $\delta$S3839(CN)$>$0 and $\delta$S3839(CN)$<$0 groups,
while $y$ is the difference between the average $\delta$CH4300 of the
$\delta$S3839(CN)$>$0 and $\delta$S3839(CN)$<$0 groups. The two centroids in the
CN-CH plane are represented by the two lage dots in each panel of
Fig.~\ref{fig_chcn}. $\Delta$CN is reported along with its uncertainty in the last
two columns of Table~\ref{tab-ratios}.}. A roughly linear trend appears when the
distance between the centroids of the CH-weak and CN-strong stars is plotted
against [Fe/H], with some sign of "saturation" at the metal-poor end, where the
indices become weak. The behaviour of M~15 clearly stands out of this simple
relation, and a prediction of the present work is that its [C/Fe] vs. [N/Fe]
anti-correlation should be significantly more extended than those of other
clusters in the sample.

\subsection{Palomar~12}

A very interesting cluster to include in our sample is Pal 12, because it is one
of the GGC traditionally associated with the Sagittarius dwarf spheroidal galaxy
(Sgr), which is at present disrupting under the strain of the Milky Way tidal
field \citep{ibata95,majewski03}. Its lower $[\alpha/\mathrm{Fe}]$ ratio
\citep{brown97,cohen04} and younger age than typical GGC
\citep{rosenberg98,marin-franch09}, its location close in Galactic longitude to
Sgr, and towards the southern extension of Sgr, its motion towards the Sgr, and
its orbital parameters indicate that it may have been captured by Sgr some 1.7~Gyr
ago \citep{dinescu+97}. 

From the chemical abundance point of view, a few [Fe/H] estimates place it in the
moderately metal-poor regime \citep[][see also
Table~\ref{int_params}]{brown97,cohen04}. In particular, \citet{cohen04} noticed
not only the lack of $\alpha$-enhancement in Pal~12, but also remarked on the
abundance of heavy elements, which is more similar to Sgr dwarf stars
\citep{bonifacio00,smecker02}. The four stars she studied showed identical
abundance pattern, with no evidence of star-to-star variations. Also
\citet{kayser08} found no sign of CH and CN anti-correlation among evolved
stars. 

With our data, a bimodal anti-correlation of CH and CN band strengths is apparent.
The group of CH-strong and CN-weak stars appears truncated and reduced in size,
probably owing to our selection (Sect.~\ref{sec-1stqc}), which excluded all stars
with S/N$<$8 in the CN spectral region. A higher S/N sample could shed some light
on this peculiar effect, which makes the CN-strong to CN-weak stars ratio appear
quite higher than in other clusters (see also Sect.~\ref{sec-trends}).

\section{Trends with cluster parameters}
\label{sec-trends}

When trying to connect the GGC properties with the presence and extension of the
anti-correlations, two parameters are usually defined: {\em (i)} the
"low-resolution community", studying the strength of molecular indices, usually
builds the ratio of CN-strong to CN-weak stars r$_{CN}$
\citep[e.g.][]{norris87,smith90,smith02,kayser08}; {\em (ii)} the "high-resolution
community" \citep[namely][]{carretta06,carretta07c} measures the extension of the
Na-O and Mg-Al anti-correlations with interquartile ranges. The two parameters
measure two different physical quantities, and both give clues to the
understanding of the anti-correlation phenomenon. In this paper we will
concentrate on the first approach, deferring the second one to a following paper,
in which we will attempt to derive [C/Fe] and [N/Fe] ratios through spectral
synthesis of the best S/N clusters.

To compare with r$_{CN}$, we selected several cluster intrinsic
(Table~\ref{int_params}) and orbital parameters (Table~\ref{ext_params}) from the
literature. Because we were looking for trends and were not particularly
interested in zeropoints and absolute quantities, we preferred papers reporting
homogeneous data, trying to avoid mixing of different sources whenever possible.

\begin{figure*}
\centering
\includegraphics[bb=50 50 470 760,clip,width=14cm]{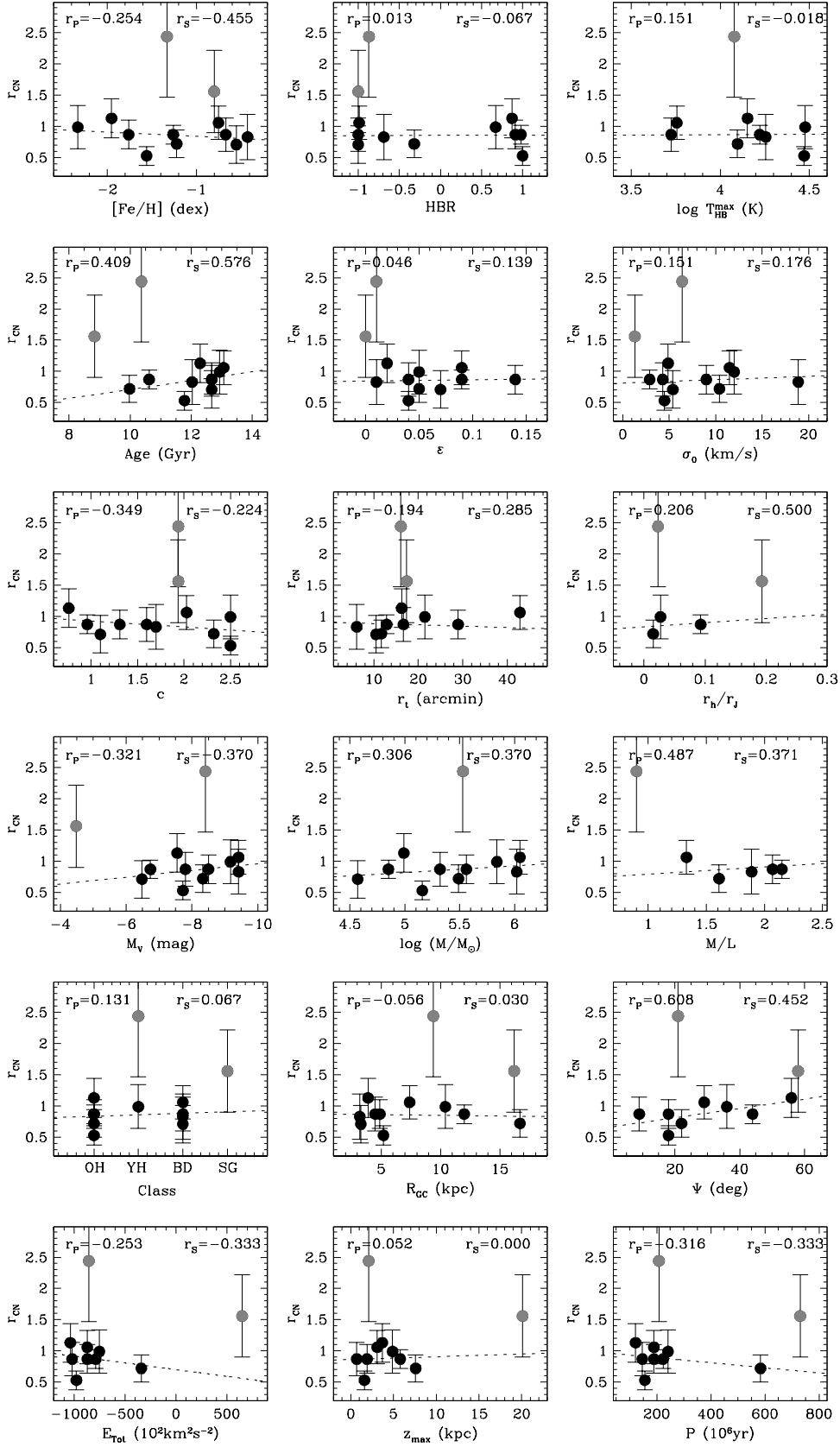}
\vspace{-0.5cm}
\caption{Run of the ratio r$_{CN}$=n$_{CN_{strong}}$/n$_{CN_{weak}}$ for our
sample clusters. In each panel, NGC~362 ans Pal~12 (grey dots) are excluded from
the relations. Dotted lines mark linear fits to the remaining clusters (black
dots) and the correlation coefficients are reported on top of each panel: r$_S$
stands for the Spearman and r$_P$ for the Pearson coefficient. For the meaning of X labels,
see Tables~\ref{int_params} and \ref{ext_params}.}
\label{fig_cnratio}
\end{figure*}

\begin{table}
\caption{Ratio of CN-strong and CN-weak stars and centroids distance.}
\label{tab-ratios}
\begin{center}
\begin{tabular}{l c c c c c c} 
\hline \hline
Cluster & n$_{CN_{strong}}$ & n$_{CN_{weak}}$ & r$_{CN}$ & $\delta$r$_{CN}$ &
$\Delta$CN & $\delta\Delta$CN \\
\hline
NGC104   & 33 & 31 & 1.06 & 0.27 & 0.209 & 0.051 \\ 
NGC288   & 60 & 69 & 0.87 & 0.15 & 0.088 & 0.023 \\ 
NGC362   & 22 &  9 & 2.44 & 0.97 & 0.063 & 0.070 \\  
NGC1851  & 18 & 25 & 0.72 & 0.22 & 0.069 & 0.026 \\  
NGC5927  & 20 & 23 & 0.87 & 0.27 & 0.244 & 0.085 \\  
NGC6352  & 10 & 14 & 0.71 & 0.30 & 0.232 & 0.075 \\  
NGC6388  & 10 & 12 & 0.83 & 0.36 & 0.270 & 0.139 \\  
NGC6656  & 26 & 30 & 0.87 & 0.23 & 0.030 & 0.031 \\  
NGC6752  & 19 & 36 & 0.53 & 0.15 & 0.063 & 0.018 \\  
NGC6809  & 27 & 24 & 1.13 & 0.31 & 0.037 & 0.025 \\  
NGC7078  & 16 & 16 & 1.00 & 0.35 & 0.080 & 0.056 \\  
Pal12	 & 14 &  9 & 1.56 & 0.66 & 0.115 & 0.063 \\  
\hline\hline
\end{tabular}
\end{center}
\end{table}

\subsection{Ratio of CN-strong to CN-weak stars} 
\label{sec-rcn}

The ratio between CN-strong and CN-weak stars tells us the relative importance of
stellar groups with different chemical composition. In the framework of the
self-enrichment scenario, for example, it becomes a fundamental constraint to
model the chemical evolution of a GGC. A correlation between r$_{CN}$ and cluster
ellipticity, defined as $\varepsilon$=1--(b/a), was initially found
\citep{norris87,smith90,smith02}, but later studies \citep{kayser08} did not
confirm it. \citet{smith90} and \citet{kayser08} found that clusters with a higher
fraction of CN-strong stars are also more luminous and therefore more massive.
This finding is supported by the trend found by \citet{smith90} with the central
velocity dispersion $\sigma_0$. While \citet{kayser08}, in their compilation of
literature estimates of r$_{CN}$, found all possible values of r$_{CN}$ between 0
and 2, \citet{car09a} and \citet{car09b}, using more homogeneus data, found that
the second -- or I as intermediate -- generation  stars (i.e., the CN-strong
stars) are between 50\% and 70\% of the total. 

We computed our r$_{CN}$ ratio by selecting stars with $\delta$S3839(CN)$>$0 for
our CN-strong group and stars with $\delta$S3839(CN)$<$0 for our CN-weak group.
The final errors on r$_{CN}$ were propagated by assuming Poissonian errors
($\sqrt{n}$) in the CN-strong and CN-weak counts\footnote{As noted in
Sect.~\ref{sec-tg}, the very small uncertainty in the placement of the
sparation line between CN-strong and CN-weak stars can have an impact on
r$_{CN}$ in the sense of moving a few stars from one group to the other. This
uncertainty ranges from the order of 1\% to a maximum of 10\%. In any case, it
is always small compared to the uncertainties quoted in Table~\ref{tab-ratios}.}
and are reported in Table~\ref{tab-ratios}. We found that even if the CH4300
region has a much higher S/N in our spectra, the spread of a corresponding
r$_{CH}$ ratio is too large to identify any correlation. We interpret this as a
consequence of the fact that CN, being a double-metal band, is more sensitive to
abundance variations and therefore r$_{CN}$ is the best indicator \citep[see
also][]{norris87}. 

The average ratio we find between the two populations, $<$r$_{CN}$$>$=
0.82$\pm$0.29, implies that the CN-strong stars -- the second generation -- tend
to constitute the minority of the total population, in contrast with the results
by \citet{car09a,car09b}. In both studies there are ample uncertainties, with our
estimated fraction of second generation stars varying from roughly 25\% to 50\% of
the total, and the same fraction estimated by \citet{car09a,car09b} varying from
about 50\% to 70\% of the total. While the discrepancy is only marginal, and
indeed the data appear consistent within the uncertainties, it is nevertheless
puzzling that these two indicators -- the CH-CN and the Na-O anti-correlations --
give somewhat different answers (see also the discussion in Sect.~\ref{sec-bimo}).

\subsection{Comparison with cluster parameters}

Our results are reported in Fig.~\ref{fig_cnratio}, where we also report on the
Spearman (r$_S$) and Pearson (r$_P$) correlation coefficients\footnote{To compute
the Spearman and Pearson coefficients and their associated probabilities, we used
the algorithms described by \citet{press92} in their {\em "Numerical recipes"}.}.
As a first note, we observe that our r$_{CN}$ values all range between 0.5 and 1.0
($<$r$_{CN}$$>$=0.82$\pm$0.29), with the only exceptions of Pal~12 and NGC~362,
which have much higher uncertainties than the rest of the clusters. Indeed, our
strict selection of spectra with S/N$>$8 could have biased our results for Pal~12
against CN-weak stars, while for NGC~362 the S/N ratio and the low number of stars
prevent us from measuring r$_{CN}$ properly (see also the errorbars in
Fig.~\ref{fig_hist}). Therefore, we have decided to leave out NGC~362 and Pal~12
from the present attempt to look for linear correlations of r$_{CN}$ and cluster
parameters. 


We do not find any highly significant correlation of cluster parameters with
r$_{CN}$. The only correlations that have significance slightly above 1\,$\sigma$,
are the ones of r$_{CN}$ with cluster age (83\%, or $\simeq$1.2\,$\sigma$),
concentration c (88\%, or $\simeq$1.3$\sigma$), total luminosity M$_V$ (82\%, or
$\simeq$1.2\,$\sigma$), total present-day mass in units of log(M/M$_{\odot}$)
(81\%, or $\simeq$1.2\,$\sigma$), orbital inclination $\Psi$ (95\%, or
$\simeq$2\,$\sigma$), and orbital period P (78\%, or $\simeq$1.1\,$\sigma$).
However weak, these relations broadly agree with past findings, except for the
correlation between r$_{CN}$ and ellipticity. In the framework of the {\em
self-enrichment} scenario, this might mean that the relative size of the two
stellar generations (or the efficiency of the second star formation episode) is
only marginally influenced by the cluster total mass and orbital parameters. The
correlation with cluster age is detected here for the first time and if confirmed,
it could mean that older clusters are more efficient in when converting gas into
the second generation of stars (but see Sect.~\ref{sec-concl}).


\section{Discussion and conclusions}
\label{sec-concl}

We can summarize the results of this paper as follows:

\begin{itemize}
\item{for the metal-rich clusters (Pal~12, 47~Tuc, NGC~5927, NGC~6352) where the
S/N of our spectra was good enough, we could detect clear anti-correlations
between the CN and CH band strengths, which were always clearly bimodal;}
\item{for the metal-poor clusters we could detect no clear anti-correlation or
bimodality, even in those cases where we knew from past studies that variations
should be present; we ascribe that to the fact that we were unfortunately not
able to gather sufficient S/N (although higher than for the metal-rich clusters)
to study CH and CN bands in these metal-poor MS stars;}
\item{the only exception was M~15, the most metal-poor of the entire sample;
which shows an abnormally large difference in band strengths for its
metallicity;}
\item{we computed r$_{CN}$, the ratio between the CN-strong and CN-weak stars,
which turned out to be on average $<$r$_{CN}>$=0.82$\pm$0.29;}
\item{no strongly significant correlation was found between r$_{CN}$ and 15
clusters parameters deduced from literature compilations;}
\item{weak correlations (slightly above 1\,$\sigma$) were found with cluster
total present-day mass and luminosity;}
\item{similarly weak correlations were found with orbital parameters, with
cluster concentration and with cluster age.} 
\end{itemize}

Now, as reviewed in Sect.~\ref{sec-intro}, the so-called {\em intrinsic}
scenarios involving various kinds of extra-mixing are not sufficient to explain
the observed chemical anomalies in GGC, although of course mixing does occur in
GGC stars \citep{gratton00}. This is especially true if one considers, as we did,
only unevolved stars that still have not undergone the so-called first dredge-up
episode. We will therefore focus on the {\em extrinsic} scenarios, and in
particular on the {\em self-enrichment} scenario.

Our first result is that {\em in all the studied clusters, those which have
sufficiently good data to show CH and CN anti-correlations among MS stars are
clearly bimodal in both the CN and CH band strengths}. This gives a very strong
support to the above scenario, especially when it predicts two (or more)
subsequent stellar generations, each polluting the gas from which the following
one forms. In this respect we can say that within our sample, we always find two
distinct groups of stars and not more. So, while exceptions are always possible
(let us not forget about, e.g., $\omega$~Cen and possibly NGC~2808), the general
tendency seems to have at most two major stellar generations in each
GGC\footnote{Of course, we cannot eclude with our study the existence of small
populations totalling less than $\simeq$10--20\% of the respective cluster
populations. Therefore, the existance of an extreme \citep[E in the notation
by][]{car09a,car09b} population is by no means excluded by our data.}.

Our second result is that {\em M~15 appears to possess exceptionally large
variations of the CH and CN band strengths, for its metallicity.} If this, as
expected, will lead to an exceptionally extended [C/Fe] vs [N/Fe]
anti-correlation, it would certainly be worthwhile to follow-up on M~15 with large
samples of high-resolution spectra, to better study the presence of multiple
populations also in the Na-O plane. Given its peculiar properties
(Sect.~\ref{sec-m15}), M~15 could even belong to the group of massive clusters
that have undergone more self-enrichment than usual, either because of their
special properties or because they could be the remnants of accreted dwarf
galaxies (such as $\omega$~Cen). Therefore, also further searches for variations
in heavier species such as Ca, Fe, or Ba should be extremely interesting.

Our third result is that {\em the ratio of CN-strong to CN-weak stars is on
average $<$r$_{CN}>$=0.82$\pm$0.29.} This challenges to some extent results by
\citet{car09a,car09b}, who find that the first or primordial (P) generation
accounts for approximately 1/3 of the total cluster population, leaving space for
a 50--70\% fraction for the second or intermediate (I) population. On the one hand
the discrepancy is -- at best -- marginal, as discussed in Sect.~\ref{sec-trends}
and given the ample uncertainties involved; on the other hand, the discrepancy
could be caused by the Na-O and the CH-CN anti-correlations sampling two sligthly
different phenomena (as discussed in Sect.~\ref{sec-bimo}). Certaintly more
studies are required to shed more light on this issue, because these kind of
discrepancies -- if confirmed -- could turn out into further clues on the
formation and evolution of GGC.

Our fourth result is that {\em no strong correlation was found between r$_{CN}$
and 15 different cluster parameters compiled from the literature.} While we must
stress that the present study is limited to 12 clusters with spectra of modest
S/N, the absence of strong corelations has one major implication in the {\em
self-enrichment scenario}: it would imply that the population ratio of the two
(or more) stellar generations is only mildly (if at all) influenced by the
cluster properties. In particular, clusters with larger present-day total masses
(and therefore luminosities) could be able to form only sligthly higher
fractions of second generation stars with respect to less massive clusters.
Similarly, the relations with cluster orbital parameters could hint that
environment has some (minor) r\^ole in determining the size of the second
stellar generation. 

In summary, the use of low-resolution spectra to study CH and CN bands in GGC
stars provides a useful complement to the high-resolution studies, needing less
observing time and data analysis effort. In particular, the use of {\em strictly
unevolved} stars. i.e., stars that have not undergone first dredge-up mixing,
and of two elements such as C and N that (unlikely Na and O) are strictly
produced in the same nuclear reaction chains and at the same temperature, can
give additional constraints to studies based on red giants only.

\begin{acknowledgements} 

We warmly thank Michael Hilker for providing the data for M~55 in electronic
form and Harald Kuntschner for an interesting discussion on index measurements.
We are grateful to Elena Valenti and Emanuele Dalessandro for their help with
photometric catalogues. We also thank Michele Bellazzini for advice on the
dynamical properties of GGC and Angela Bragaglia for interesting discussions on
chemical anomalies in GGC. EP acknowledges support by the ESO Visitorship
programme both in Germany and in Chile. RC acknowledges the funds by the Spanish
Ministry of Science and Technology under the MEC/Fullbright postdoctoral
fellowship program. MZ acknowledges support by the FONDAP Center for
Astrophysics 15010003, BASAL CATA PFB-06, Fondecyt Regular 1085278 and
MIDEPLAN's Milky Way Millennium Nucleus P07-021-F. 

\end{acknowledgements}

\end{document}